\documentclass{article}

\usepackage{arxiv}

\usepackage[utf8]{inputenc} % allow utf-8 input
\usepackage[T1]{fontenc}    % use 8-bit T1 fonts
\usepackage[hidelinks=true]{hyperref}       % hyperlinks
\usepackage{url}            % simple URL typesetting
\usepackage{booktabs}       % professional-quality tables
\usepackage{amsfonts}       % blackboard math symbols
\usepackage{microtype}      % microtypography
\usepackage{graphicx}
\usepackage[round]{natbib}
\usepackage{amsmath}

\usepackage{graphicx}% Include figure files
\usepackage{dcolumn}% Align table columns on decimal point
\usepackage{bm}% bold math

\usepackage{multirow}
\usepackage{amssymb}% http://ctan.org/pkg/amssymb
\usepackage{pifont}% http://ctan.org/pkg/pifont
\newcommand{\cmark}{\ding{51}}%
\newcommand{\xmark}{\ding{55}}%
\usepackage[utf8]{inputenc}
\usepackage[T1]{fontenc}
\usepackage{mathptmx}
\usepackage{etoolbox}
\usepackage{xcolor}
\usepackage{soul} 
\usepackage{siunitx}

\usepackage[font=small,labelfont=bf]{caption}

\newcommand\myhl[1]{\textcolor{black}{#1}}
\newcommand{\hlwhite}[1]{{\sethlcolor{white}\hl{#1}}}

\title{Turbulent Injection assisted by Diffusion Models for Scale Resolving Simulations}

\author{ 
    Margaux~Boxho \\
	Département HiFi CFD \& CAA\\
    Cenaero, Rue des Frères Wright 29, \\
    6041 Charleroi, Belgium\\
	\texttt{margaux.boxho@cenaero.be} \\
	\And
    Joachim~Dominique \\
	Département Turbomachines\\
    Cenaero, Rue des Frères Wright 29, \\
    6041 Charleroi, Belgium\\
	\texttt{joachim.dominique@cenaero.be} \\
    \And
    Tariq~Benamara \\
	Département Machine Learning \& Optimisation\\
    Cenaero, Rue des Frères Wright 29, \\
    6041 Charleroi, Belgium\\
	\texttt{tariq.benamara@cenaero.be} \\
    \And
    Michel~Rasquin \\
	Département HiFi CFD \& CAA\\
    Cenaero, Rue des Frères Wright 29, \\
    6041 Charleroi, Belgium\\
	\texttt{michel.rasquin@cenaero.be} \\
    \And
    Lionel~Salesses \\
	Département Machine Learning \& Optimisation\\
    Cenaero, Rue des Frères Wright 29, \\
    6041 Charleroi, Belgium\\
	\texttt{lionel.salesses@cenaero.be} \\
    \And
    Caroline~Sainvitu \\
	Département Machine Learning \& Optimisation\\
    Cenaero, Rue des Frères Wright 29, \\
    6041 Charleroi, Belgium\\
	\texttt{caroline.sainvitu@cenaero.be} \\
    \And
    Gilles~Louppe \\
	Département d'électricité, électronique et informatique,\\
    University of Liège, Allée de la Découverte 10, \\
    4000 Liège, Belgium\\
	\texttt{g.louppe@uliege.be} \\
    \And
    Thomas~Toulorge \\
	Département HiFi CFD \& CAA\\
    Cenaero, Rue des Frères Wright 29, \\
    6041 Charleroi, Belgium\\
	\texttt{thomas.toulorge@cenaero.be} \\
}

\date{}

\renewcommand{\shorttitle}{Diffusion Model for Turbulence Injection}

\hypersetup{
pdftitle={Turbulent Injection assisted by Diffusion Models for Scale Resolving Simulations},
pdfsubject={q-bio.NC, q-bio.QM},
pdfauthor={M.~Boxho, J.~Dominique, T.~Benarama, M.~Rasquin, L.~Salesses, C.~Sainvitu, G.~Louppe, T.~Toulorge},
pdfkeywords={Diffusion Models, Decaying Homogeneous Isotropic Turbulence, Turbulence Simulations, Scale Resolving Simulations, Deep Learning},
}

\begin{document}
\maketitle
\vspace{-8mm}
\begin{abstract}
The present research proposes a new memory-efficient method using diffusion models to inject turbulent inflow conditions into Large Eddy Simulation (LES) and Direct Numerical Simulation (DNS) for various flow problems. A guided diffusion model was trained on Decaying Homogeneous Isotropic Turbulence (DHIT) samples, characterized by different turbulent kinetic energy levels and integral length scales. Samples generated by the diffusion model accurately reproduce turbulence statistics, such as the energy spectrum and the two-point autocorrelation functions, while preserving the ability to generate instantaneous three-dimensional velocity fields with detailed fluctuations. Physical representativeness is also evaluated by injecting the \textit{synthetic} samples into a free domain  (i.e., without any wall boundary) through an inlet boundary condition. The method demonstrates promising results regarding energy spectrum, spatial correlation, turbulent kinetic energy level, and integral length scales without increasing the development distance as compared to a library-based method. The following article has been submitted to/accepted by Physics of Fluid (AIP Publishing LLC). After publication, it will be available \url{https://doi.org/10.1063/5.0278541}. 
\end{abstract}

% keywords can be removed

% ---
% Literature review, existing methods, objectives
% ---

\section{\label{sec:intro}Introduction}

The enhancement of computational capabilities and the increased accessibility of large-scale clusters have facilitated the use of Large Eddy Simulation (LES) and Direct Numerical Simulation (DNS) for real compressible flow configurations. To ensure realistic simulations, high-quality and highly representative boundary conditions are essential. One of the main challenges is reproducing, at the domain inlet, an unsteady turbulent flow compliant with experimental conditions such as turbulence grids~\citep{CNCNTZM2011}. Since such an inflow varies stochastically and continuously in space and time, defining an ideal turbulent inflow remains a complex task. The fluctuations must mimic real turbulence, as they will impact the downstream flow dynamics~\citep{SGU2024}. The constructed turbulent flow field should satisfy statistical turbulent characteristics in both space and time, be fully developed as quickly as possible~\citep{DBL2015}, and should be divergence-free to avoid the injection of spurious pressure waves (i.e., acoustic effects).

\subsection{State of the art}

A variety of methods have been proposed to obtain high-quality, fully turbulent inflow data, which can be categorized into four groups: (i)~transition-inducing methods,  (ii)~recycling-rescaling methods, (iii)~synthetic inflow generators, and (iv)~turbulence library-based methods.

%The transition-inducing method is one of the most natural approaches to trigger the transition of laminar-to-turbulent boundary layers. In its original version, the inlet, which refers to a plane where the flow is laminar, is positioned far upstream from the region of interest to allow the boundary layer to evolve naturally into a turbulent state. By construction, the computational domain needs to be very long to allow for this natural transition, and this constraint drastically increases the cost of the simulation. Artificial (numerical~\citep{SO2012,PBCR2019} or geometric~\citep{CPLP2022}) disturbances can be introduced to accelerate the transition. These numerical perturbations should be handled carefully, as they can produce spurious fluctuations~\citep{M2012}. % LONG VERSION 

The transition-inducing method requires a very long extended domain entry to allow for the natural transition of the Boundary Layer (BL) to a turbulent state, which drastically increases the computational cost. The introduction of artificial~\citep{SO2012,PBCR2019,CPLP2022} perturbations to accelerate the transition should be handled carefully, as they can produce undesired fluctuations~\citep{M2012}.

%The Recycling-Rescaling Method (RRM) was initially proposed by Lund and Wu~\citep{LWS1998} for spatially developing incompressible boundary layers. The recycling method strategy operates as follows: a plane of instantaneous velocities located inside the computational domain is extracted and reinjected at the inlet. This approach allows the flow to develop continuously and ultimately generate turbulence. \citet{W17} classified the recycling method into the strong recycling method, where a strict periodic condition is applied, and the weak recycling method, where the data is recycled and rescaled to satisfy statistics before being imposed at the inlet. RRM for compressible flows has been proposed by~\citet{XM04}, where the temperature is coupled to the velocity based on Morkovin's hypothesis. The main drawback of the RRM is the strong spatial correlation imposed between the inlet and extraction position. %LONG VERSION

The Recycling-Rescaling Method (RRM), initially proposed by~\citet{LWS1998} for spatially developing incompressible BLs, extracts a plane of instantaneous velocities located inside the computational domain and reinjects it at the inlet, with a potential \textit{a priori} scaling allowing for statistics matching before being imposed at the inlet~\citep{W17,XM04}. The main drawback of the RRM is the strong spatial correlation imposed between the inlet and extraction positions.

%For complex geometries or industrial applications, setting up auxiliary periodic boundaries can be challenging, and no library data sources may be available. In such cases, synthetic inflow generators are the preferred method. Synthetic methods do not simply impose random perturbations on a mean profile at the inlet of the domain, which is insufficient to induce turbulence and may generate acoustic waves. Compared to precursor simulations, they alleviate the issues related to computational cost and trial-and-error methodology by constructing turbulence-like fluctuations from uncorrelated data and known target properties (e.g., the Reynolds stresses, the integral length scales, and the turbulent kinetic energy, to cite a few). Synthetic inflow generators are distinguished primarily by how they enforce spatial and temporal coherence in the inflow signal. They can be subdivided into six main strategies: (a)~the spectral-representation-based approach~\citep{LLM1992,SSC01,BGC04,DB06,D08}, (b)~the Proper Orthogonal Decomposition (POD) approach~\citep{DLDB1999,BHL1993}, (c)~the digital filter approach~\citep{KSJ03,MKJJ06,HNF11,DBL18,TSSGP20,HHCdM22,DHdM23}, (d)~the volumetric-forcing-based approach~\citep{SB17}, (e)~the synthetic eddy approach~\citep{S02,JBLP06,PWGDS09,XGBSL18}, and (f)~the cell perturbation methods~\citep{MEKvBM15,DLUS19}. % LONG VERSION

Synthetic inflow generators are used to construct turbulence-like fluctuations from uncorrelated data and known target properties. They differ in how they enforce spatial and temporal coherence in the inflow signal, and are classified into six main strategies: (a)~the spectral-representation-based approach~\citep{LLM1992,SSC01,BGC04,DB06,D08}, (b)~the Proper Orthogonal Decomposition approach~\citep{DLDB1999,BHL1993}, (c)~the digital filter approach~\citep{KSJ03,MKJJ06,HNF11,DBL18,TSSGP20,HHCdM22,DHdM23}, (d)~the volumetric-forcing-based approach~\citep{SB17}, (e)~the synthetic eddy approach~\citep{S02,JBLP06,PWGDS09,XGBSL18}, and (f)~the cell perturbation methods~\citep{MEKvBM15,DLUS19}. Despite their relatively low computational cost, synthetic methods still rely on flow information that is difficult to obtain, especially for complex flow configurations where the Reynolds stress components are needed to determine the turbulence anisotropy. Although the fluctuations generated by synthetic methods mimic turbulent structures and produce realistic temporal and spatial correlations using analytical expressions, these fluctuations are not fully consistent with real turbulence~\citep{KP06} and, require an extension of the computational domain to allow the development of more realistic turbulent flows. Higher-order statistics are matched after a certain development distance from the inlet boundary. 

Turbulence library-based methods, also named precursor methods, rely on an auxiliary computation, which can be run \textit{a priori} or concurrently to generate turbulent data. This data is then imposed at the inlet of the main simulation through a boundary condition. Such a method provides a high-quality and realistic inflow~\citep{RTBH2023}, as minimal deviations from the actual configuration are used to generate it. However, this realism comes at the cost of additional computational and memory requirements. In addition, the method requires adjustment of the precursor simulation parameters, with trials-and-errors of both the main domain and the precursor simulations to obtain the appropriate turbulence statistics at the target location in the main computational domain. This back-and-forth process can be time-consuming, adds complexity to the workflow, and requires significant computational and human resources. 

\subsection{Our contribution}

The purpose of the present work is to develop a novel turbulence injection technique that can generate a field of realistic turbulent fluctuations from a limited set of parameters. The first objective is to streamline the trial-and-error process required to establish the appropriate turbulence statistics at a specific location within the main computational domain by eliminating the need for an additional DHIT simulation at each new Reynolds number. Our second objective is to reduce the memory consumption of the turbulence library-based method while maintaining spectrally realistic turbulence~\citep{KPBK01}. In addition, the fluctuations should be established within the shortest distance possible from the inlet, making that distance a reliable metric for evaluating the efficiency of a method and the validity of the underlying assumptions. The third objective is to obtain physical representativeness comparable to library-based and transition-inducing methods, while maintaining a low computational cost and featuring easy parametrization similar to synthetic methods.

%The purpose of the present work is to develop a novel turbulence injection technique that can generate a field of realistic turbulent fluctuations from a limited set of parameters. The first objective is to reduce the trial-and-error runs to set up the appropriate turbulence statistics at a given location in the main computational domain, as well as the memory consumption of the turbulence library-based method, while maintaining spectrally realistic turbulence~\citep{KPBK01}. The fluctuations should be established within the shortest distance possible from the inlet, making that distance a reliable metric for evaluating the efficiency of a method and the validity of the underlying assumptions. The second objective is to obtain physical representativeness comparable to library-based and transition-inducing methods, while maintaining a low computational cost and featuring easy parametrization similar to synthetic methods. 

Recent advances in deep learning algorithms, coupled with the increasing power of GPUs and the generation of high-fidelity data, have led to the exploration of novel data-driven approaches to improve existing turbulence injection methods or to develop new ones. To this end, we use state-of-the-art Diffusion Models~\citep{HJA2020} (DM), a class of latent variable models inspired by non-equilibrium thermodynamics, to generate three-dimensional triply periodic boxes of DHIT. 

Along with Generative Adversarial Networks (GANs) and Variational Auto-Encoders (VAEs), DMs belong to the class of generative models that, by learning the probability distribution of the training data, can produce novel outputs that retain the statistical properties of the original data.  All generative models face a fundamental trilemma: balancing sample quality, generation speed, and mode coverage. GANs produce high-quality samples efficiently but often suffer from mode collapse, limiting their ability to capture the full data distribution. In contrast, VAEs and normalizing flows provide better mode coverage but typically generate lower-quality samples~\citep{XKV2021}. Finally, DMs are high-quality image generator that outperforms GANs~\citep{DN2024} and achieve strong mode coverage. However, their inference process is relatively slow due to the iterative denoising steps required for generation. To mitigate this, ongoing research explores accelerating inference using Stochastic Differential Equation (SDE) solvers~\citep{YSDKKEP2021,ZC2023}. Despite slower sampling, DMs are easier to define and more efficient to train than GANs. DMs are thus a promising tool to develop a novel turbulence injection method suitable for various flow problems and capable of producing high-quality samples. 

Although the present work is restricted to homogeneous isotropic freestream turbulence, we aim to develop a method that can be extended to other types of turbulent inflows (e.g., anisotropic turbulence, boundary layers, and wakes).

\subsection{Related works}

\citet{FNKF2019} trained an Auto-Encoder (AE) on two-dimensional snapshots of velocity and pressure fluctuations of a turbulent channel flow at $Re_\tau=180$. \textit{A posteriori}, the model provided the time-dependent inflow condition for the DNS of a turbulent channel flow with proper inflow-outflow conditions and successfully maintained turbulence. 

\citet{KL20} used a GAN coupled with Recurrent Neural Networks (RNN) to predict time series of two-dimensional snapshots. The model was trained on turbulent channel flows at multiple Reynolds numbers ($Re_\tau=\{180, 360,540\}$). The RNN-GAN produced a statistically stationary flow with small-scale turbulent structures near the wall, closer to DNS than AE predictions~\citep{FNKF2019}.

\citet{KNTS2021} trained convolutional VAE coupled to GAN on two-dimensional snapshots of forced isotropic turbulence simulations to generate natural-looking fluid flows. The model produced highly detailed predictions but suffered from artifact generation.  

\citet{YZYVL22} proposed a combination of a transformer and MS-ESRGAN (a specific implementation of GAN) to generate turbulent inflow conditions for spatially developing Turbulent Boundary Layer simulations. Reasonable agreement with DNS was observed at Reynolds numbers used to train the network. The model also showed good capabilities to reproduce turbulent statistics, such as the velocity components spectra.

\citet{LPFDWCW2024} recently proposed CoNFiLD-inlet, a model that combines a stable diffusion model with a conditional neural field-encoded latent space to generate time series of 2D turbulent inlet conditions for high-fidelity turbulent channel flow simulations. The model generalizes across various Reynolds numbers by initializing the flow predictions based solely on the Reynolds number and the long-term simulation horizon is obtained using an auto-regressive process in the latent space. While their approach shares similarities with the present work, key differences exist in both the training data and the methodology used for time series prediction. Their study focuses on turbulent channel flow inlets, whereas this paper examines DHIT. Additionally, their method predicts time series using an auto-regressive scheme in a latent space obtained over a neural field, which requires joint optimization of both the encoder parameters and the latent representation. In contrast, our approach constructs long-term predictions by directly generating 3D turbulence boxes.

In the context of time series prediction, \citet{TKK2024} developed a methodology that uses orthogonal wavelet analysis, long short-term memory (LSTM), and three-dimensional convolutional neural networks (3D CNN) to predict the time evolution of 3D HIT. Wavelets are an efficient tool for representing multiscale and intermittent fields, and their projections reduce the size of the input data and memory consumption during training.

\citet{LLHPG2024} proposed developing a generative method that can directly learn the manifold of all possible turbulent states and act as a surrogate model for replacing numerical flow solvers. The authors trained a DDPM and included Dirichlet boundary conditions to guide the diffusion process. Nonetheless, the proposed method could not generate sequences, limiting the analysis of dynamical phenomena. To overcome this weakness, \citet{SLG2024} used 4D generative diffusion model combined with physics-informed guidance for the generation of realistic sequence of 3D flows. Their architecture is a UNet composed of bi-directional ConvGRU layers.

\subsection{Paper organization}

The paper is organized as follows. Section~\ref{sec:data} presents the database consisting of several DHIT samples characterized by different Turbulent Kinetic Energy (TKE) levels and integral length scales $L_{int}$. In contrast to the models presented in the literature review, the proposed model directly predicts the three-dimensional velocity fields instead of decoupling space and time. Section~\ref{sec:DDPM} introduces the two trained models: (1) an unconditional Diffusion Model (DM), and (2) a Classifier-Free (CF) DM, which allows the network to be conditioned on different parameters.  Section~\ref{sec:res} is entirely dedicated to the CF-DM and its \textit{a priori} and \textit{a posteriori} assessment on multiple DHITs. Section~\ref{sec:conclu} concludes the paper with perspectives to further improve the proposed data-driven method.

% Section~\ref{sec:valid} focuses on the unconditional DM trained on a single DHIT box and serves to validate the metrics indicating that the turbulence generated by a diffusion model is realistic.

% ---
% Deep generative models and score-based model
% Database generation and non-dimensionalization
% Argo-DG and implementation of the boundary condition
% ---
\section{\label{sec:data}Data}

Section~\ref{subsec:solver} introduces the in-house flow solver, Argo-DG, used to generate the database. Section~\ref{subsec:Precursor} describes the injection of turbulent fluctuations based on the precursor method through a boundary condition. Section~\ref{subsec:data} introduces the database consisting of several DHIT simulations and defines the statistics of interest, which will serve as metrics to evaluate the physical representativeness of the synthetic samples.

\subsection{\label{subsec:solver}Argo-DG Flow Solver}

The in-house flow solver, Argo-DG, developed at Cenaero, is a high-order Discontinuous Galerkin (DG) flow solver for the compressible Navier-Stokes equations. \citet{H2013} implements the Discontinuous Galerkin method using the Symmetric Interior Penalty (SIP) technique for the discretization of diffusive terms. DG methods are a specific class of the finite element method, where the functional space allows for discontinuities in the solution at the interfaces between elements. At these interfaces, a Riemann solver treats the convective terms, while the SIP method controls the diffusive terms. Using appropriate polynomial approximation spaces, the method features high-order accuracy, including low dissipation and dispersion properties. The compact formulation lends itself to efficient implementation on high performance computing systems: Argo-DG has demonstrated very high scalability (up to 100,000's of cores) through a hybrid parallelism model that incorporates both Message Passing Interface and Open Multi-Processing. These characteristics are particularly well-suited for large scale-resolving CFD simulations.

For time discretization, the solver uses the second-order accuracy Backward Difference Formula (BDF2). This implicit time-stepping method is widely used due to its attractive stability. At each time step, a nonlinear problem resulting from this implicit integration is solved using a Newton/GMRES method, preconditioned with an element-wise block-Jacobi technique. 

Concerning turbulence modeling, Argo-DG implements the implicit LES approach. The numerical dissipation of the underlying high-order DG scheme spectrally acts similarly to the explicit subgrid-scale models traditionally used in LES~\citep{CHBW2014}.

\subsection{\label{subsec:Precursor}Precursor method in Argo-DG}

\begin{figure*}[h!t]
    \centering
    \includegraphics[width=\linewidth]{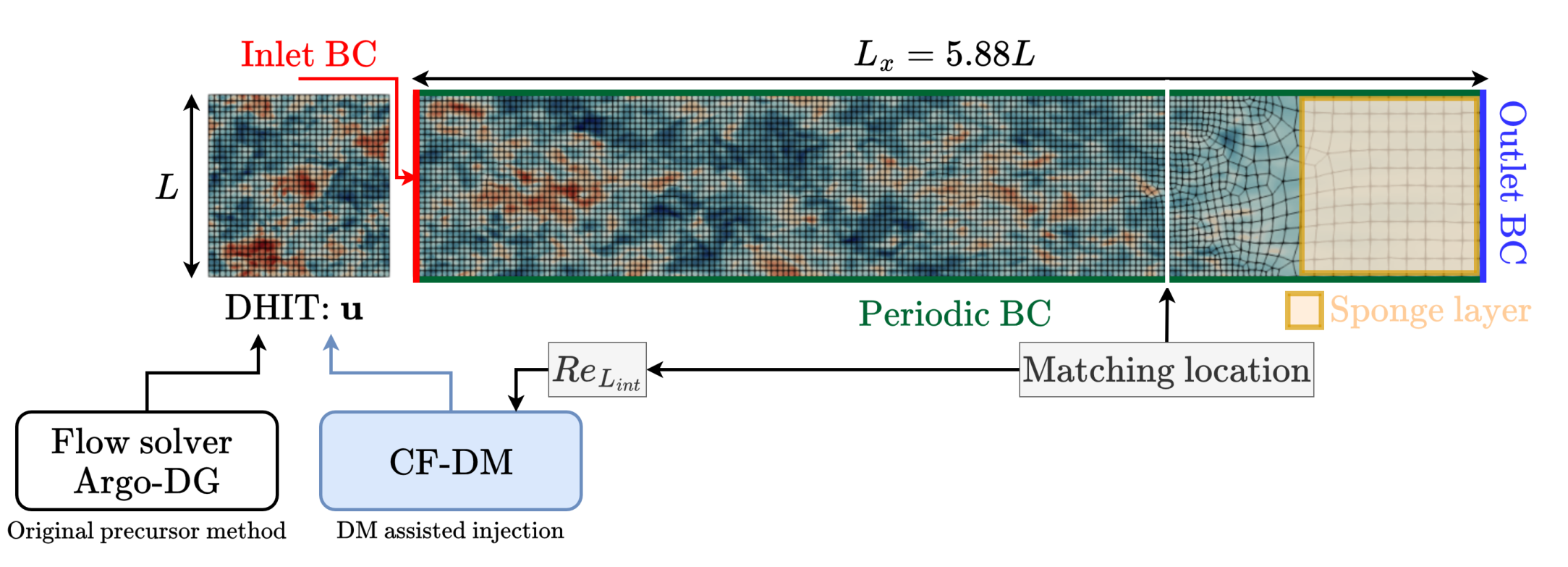}
    \caption{Sketch of the domain in the $x-y$ plane, comprising the proper boundary conditions, the mesh, the location of the sponge layer, the matching location at which the experimental data needs to be matched, and the trial-and-error loop to define the appropriate conditioning ($Re_{L_{int}}$) of the diffusion model for the generation of a novel turbulent inlet velocity field $\bm{u}$.}
    \label{fig:framework}
\end{figure*}

The library-based method is implemented in Argo-DG as a boundary condition. The velocity fluctuations $\bm{u}^{\prime}$ obtained from the precursor simulation are combined with the mean flow $\overline{\bm{u}}_{in}$ at the inlet of the main simulation \myhl{(see Figure~\ref{fig:framework})}. This technique makes two main assumptions: (i) the Taylor frozen turbulence hypothesis and (ii) the spatial homogeneity hypothesis. The former assumes that the turbulence time scales are much larger than the convection time scales and $\vert \bm{u}^{\prime}\vert \ll \vert \overline{\bm{u}}_{in} \vert$ to neglect the nonlinear interaction between turbulence and mean advection. The latter assumes that the flow is spatially homogeneous, even in the flow direction.

Under these hypotheses, the precursor solution, frozen at a time corresponding to the desired $\mbox{TKE}$ and integral length scale $L_{int}$, is sufficient. The $\mbox{TKE}$, denoted $k$, is computed as
\begin{equation}
    k = \dfrac{1}{2}\left( u^\prime u^\prime + v^\prime v^\prime + w^\prime w^\prime\right),
    \label{eq:tke}
\end{equation}
where $(u^\prime,v^\prime,w^\prime)$ are the three components of velocity fluctuations in the $(x,y,z)$-directions, respectively. The velocity fluctuations are interpolated onto the inlet plane which is swept at the main inlet velocity through the precursor solution. These interpolated fluctuations are added to the mean velocity yielding the full turbulent inlet velocity $\overline{\bm{u}}_{in} + \bm{u}^{\prime}$. From this turbulent inlet velocity, along with the imposed mean total pressure and mean total temperature, the conservative variables are computed and used to impose the boundary condition through numerical fluxes as usually done in DG. 

However, due to computational constraints on the domain size, it is often impossible to reproduce in numerical simulations the inlet turbulence length scales found in experiments. Moreover, high inlet turbulence intensity can induce nonlinear effects also impacting the dissipation rate in the domain. These difficulties can be mitigated by estimating the turbulence intensity that exists near the stagnation point on the experimental test article, and matching numerically the turbulence characteristics at that point instead of the inlet plane. This strategy is particularly cumbersome to implement with the library-based method, as it requires trial-and-error computations of both precursor and main simulations.

\myhl{As sketched in Figure~\ref{fig:framework}, the present objective is to accelerate the generation of the turbulent inlet velocity field by eliminating the need for a flow solver. A classifier-free diffusion model, conditioned to a Reynolds number, that encapsulates the level of TKE and the integral length scale, is trained to reproduce a realistic turbulent field at every Reynolds number within the training limits. In a trial-and-error loop to determine the most appropriate inlet boundary condition to properly match the experiments, the interpolation capability of the diffusion model coupled with a fast sampler is crucial.}

\subsection{\label{subsec:data}Decaying Homogeneous Isotropic Turbulence}

In wind tunnel experiments, precise control of the inflow turbulence is essential for diverse applications, including accurately replicating operating conditions in compressor and turbine flows~\citep{LN1976}, ensuring reliable acoustic measurements by minimizing spurious contributions from turbulent inflow~\citep{SAPJGG2019}, or generating atmospheric boundary layer inflow. Various turbulence control methods exist, often categorized as passive or active techniques. Passive strategies, such as bellmouth inlets, grid screens, honeycomb turbulence grids, and wire meshes, can be adapted to shape specific inflow profiles~\citep{GLBLGDDEMF2024}. Active techniques, including air jets, vortex generators, and fan arrays, provide dynamic control of turbulence characteristics~\citep{CB2023}.

Among these methods, turbulence grids are widely used to to generate isotropic freestream turbulence in front of blades (e.g., low-pressure turbine cascades~\citep{RDN2015}). Experimentalists can modify the grids (e.g., horizontal bar spacing, bar shape, grid to leading edge distance) to change both the level of Turbulence Intensity ($\mbox{TI}\%$), expressed as the ratio $\vert \bm{u}^\prime\vert / \vert \overline{\bm{u}}_{in} \vert$, and $L_{int}$. To numerically replicate these experiments using turbulence library-based methods, the canonical case of DHIT~\citep{Pope2000}, is considered. The DHIT is carried out in a three-dimensional periodic cube of size $L$, where $L$ is set to the span of the cascade blade simulation. 

The simulation is initialized with a Passot-Pouquet spectrum~\citep{PP1987}, distributing a given amount of energy into several modes. This initialization in Fourier space is then converted into physical space (i.e., fluctuations of the velocity field) using the Rogallo method~\citep{R1981}. 

The spatial resolution of the box is $(n_x,n_y,n_z) = (32\times 32 \times 32)$ combined with a fourth-order polynomial approximation, resulting in an effective resolution of $(128\times 128 \times 128)$. An implicit time scheme is used with a time step of $6.25\,\cdot\,10^{-7}$ [s], resulting in a Courant-Friedrichs-Lewy (CFL) condition of $0.25$ at the onset of the simulation. The CFL will steadily decrease with the energy during the simulation.

\subsubsection{Databases}

\begin{figure*}[h!t]
    \centering
    \includegraphics[width=.49\linewidth]{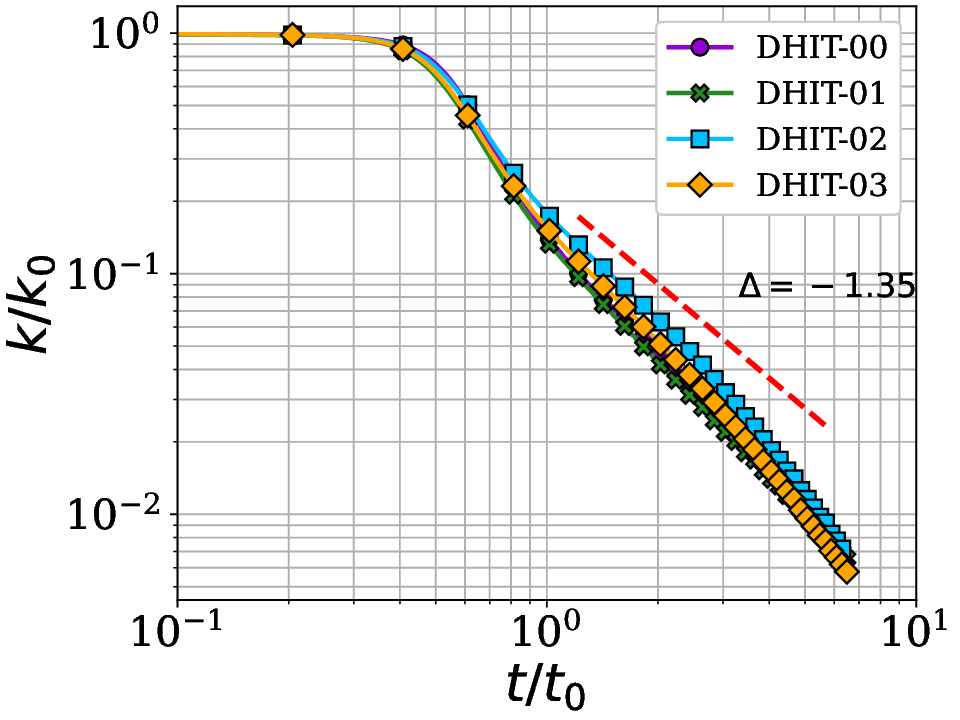}
    \includegraphics[width=.49\linewidth]{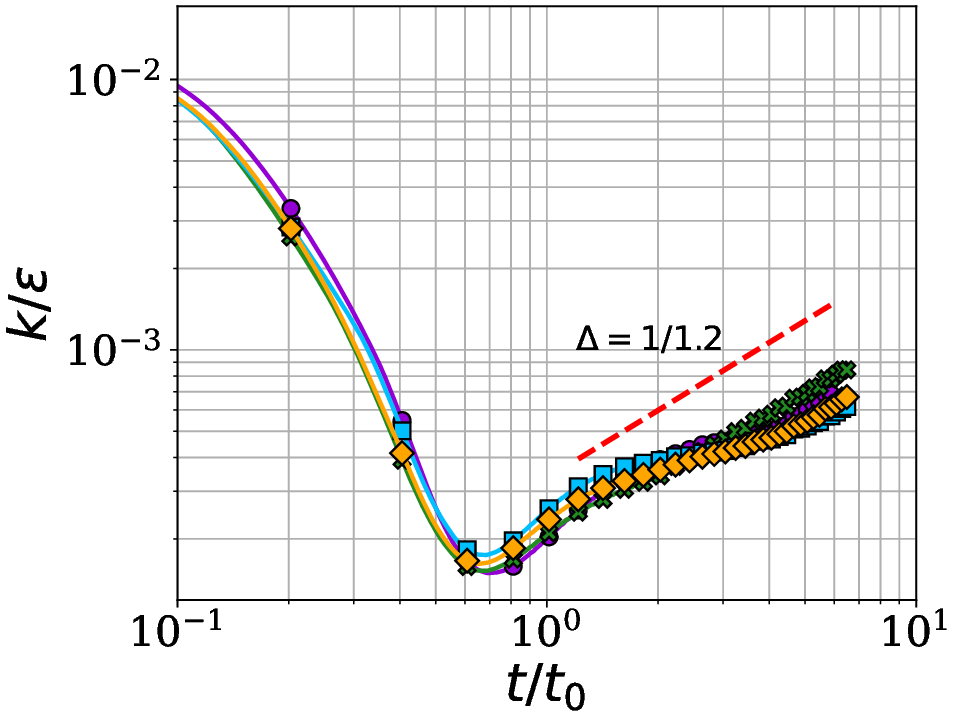}
    \caption{\label{fig:database:TKE:epsilon}On the left, the evolution of the TKE; on the right, the evolution of the ratio between the TKE and the dissipation. The dashed red lines indicate the theoretical curves.}
\end{figure*}

The time at which the DHIT is frozen is usually based on the distance between the inlet plane and the object of interest (in the computational domain) and the pair $(\mbox{TI},L_{int})$ given by the experimentalist, so that the DHIT can reach the desired statistical properties in the vicinity of the stagnation point on the object. TI can be expressed in terms of the turbulent kinetic energy and assuming, $u^\prime \approx v^\prime \approx w^\prime$, 
\begin{equation}
    \mbox{TI} = \dfrac{\sqrt{\frac{2}{3}k}}{\vert \overline{u}_{in} \vert},
\end{equation} 

Four DHITs initialized with the same Passot-Pouquet spectrum were simulated with a different random seed. They lead to different instantaneous snapshots while having similar turbulent properties. A total of 214 (TKE, $L_{int}$)-related boxes were extracted from these simulations. The TKE level ranges from $12.02$ to $363.23$, while the normalized integral length scale ($L_{int}/L$) ranges from $0.114$ to $0.181$. The frozen DHITs must be extracted when the turbulent cascade is established, which corresponds to the alignment of the evolution of (1) the non-dimensionalized TKE, and (2) the ratio between the TKE and the dissipation $\epsilon = -\frac{\mathrm{d}k}{\mathrm{d}t}$ in a log-log plot, with the theoretical targets as shown in Figure~\ref{fig:database:TKE:epsilon}.

\begin{figure*}[h!t]
    \centering
    \includegraphics[width=\linewidth, trim=0 220 210 0, clip=true]{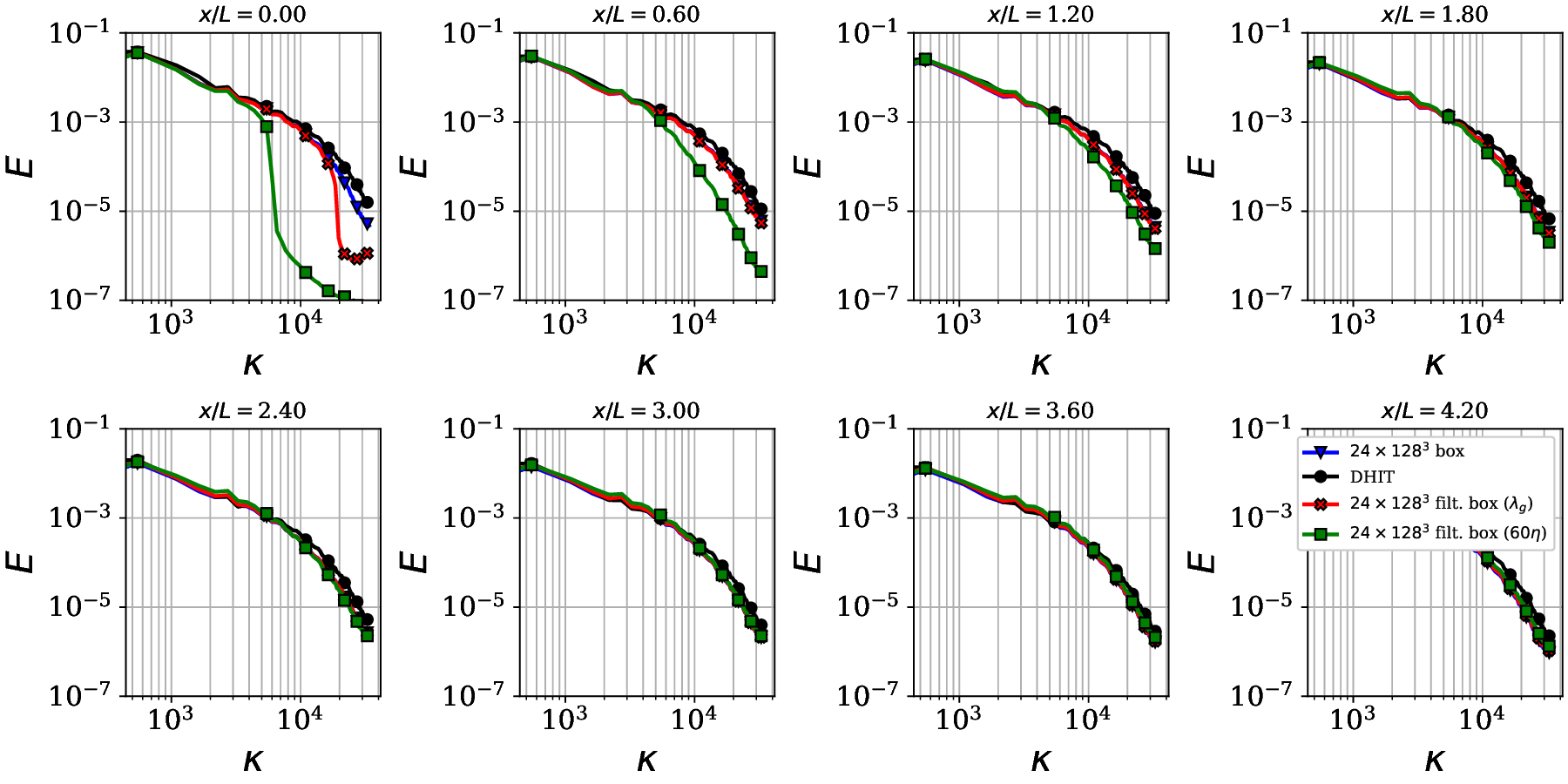}
    \caption{\textbf{Freestream turbulence injection at TI$\bm{=15.2\%}$} 2D energy spectrum at three streamwise locations $x/L=\{0,0.6,1.2\}$ in the free domain. Black circles corresponds to the original DHIT. Precursor method using DHIT boxes of size $128^3$, $64^3$, and $32^3$ are drawn in blue triangles, red crosses, and green squares, respectively. }
    \label{fig:one:filtered:spectrum}
\end{figure*}

\paragraph{Resolution} After conducting several numerical experiments, the optimal resolution for the box has been set to \(64 \times 64 \times 64\). This resolution effectively resolves the integral length scale up to the limit between the inertial and dissipation ranges defined as the Taylor microscale~\citep{Pope2000} ($\lambda_g$). It avoids wasting resources trying to reproduce all the small scales down to the Kolmogorov scale ($\eta$), as these can be restored by injecting the DHIT into a sufficiently refined domain. An accurate flow solver will naturally restore these scales based on the principles of turbulent cascade. \myhl{Figure~\ref{fig:one:filtered:spectrum} shows the evolution of the two-dimensional energy spectrum of the turbulence injected by the original precursor method using DHIT boxes of sizes $128^3$, $64^3$, and $32^3$. After less than one box length, the $64^3$ turbulence has a similar energy spectrum to the $128^3$, indicating that the smallest scales are naturally restored by the Argo-DG flow solver. The $32^3$ turbulence has a much different spectral content. Therefore, this resolution is insufficient to properly restore all turbulent scales.}

\paragraph{Non-dimensionalization} To deal with several DHIT boxes at various $\mbox{TKE}$ levels, the velocity fluctuations are scaled as a Reynolds number using the corresponding integral length scale $L_{int}$. This length scale is computed as, 
\begin{equation}
    L_{int} = \dfrac{1}{3}\left(L_{00} + L_{11} + L_{22}\right),
    \label{eq:lint}
\end{equation}
where $L_{ii}$ corresponds to the integral length scale of the $i^{th}$ velocity component evaluated along the $i$-direction. Due to isotropy, $L_{00}\approx L_{11}\approx L_{22}$. The velocity fluctuations are thus scaled as follows, 
\begin{equation}
    \bm{u}^{\prime \ast} = \dfrac{\bm{u}^\prime L_{int}}{\nu},
\end{equation}
where $\nu$ is the kinematic viscosity. 

\paragraph{Data augmentation} To improve the robustness of the neural network training process, data augmentation is applied to expand the database size. As the DHIT is periodic, translation in each direction satisfy the NS equations, making them suitable for data augmentation. Since the 3D box is isotropically discretized with $64$ points in each direction, $64^3$ translations can be applied to each $214$ available DHIT. This results in $56{,}098{,}816$ boxes in the database. However, the translation of one pixel does not produce a very different field, resulting in a high correlation between boxes. 

\subsubsection{Statistics of interest}

The statistics of interest assess the capability of the model to generate realistic turbulent fields, and they must answer the questions: \textit{"Are the DM able to generate descent synthetic turbulence~?"} In the present research, four statistics are considered: the energy spectrum, the two-point autocorrelation functions~\citep{Pope2000}, the Barycentric triangle, and the vorticity distribution. The level of TKE can be extracted by integrating the energy spectrum along the wavelength. The integral length scale can be extracted from the integration of the longitudinal and transverse correlation functions. These four statistical quantities are necessary to characterize the DHIT. 

\paragraph{Energy spectrum} The energy spectrum quantifies how the TKE is distributed between the different vortex sizes. The energy spectrum can be subdivided into three distinct regions, describing the energy cascade as initially explained by~\citet{R2007} and later by~\citet{K1991}: (i) the energy-containing region, (ii) the inertial region, and (iii) the dissipation region. 

\paragraph{Two-point autocorrelation functions} Following~\citet{Pope2000}, the two-point autocorrelation function is defined for an homogeneous and isotropic turbulence, with zero mean and velocity fluctuation $u^\prime(t)$, as, 
\begin{equation}
    R_{ij}(\bm{r},t) = \langle u_i(\bm{x}+\bm{r},t), u_j(\bm{x},t)\rangle, 
\end{equation}
where $i,j=\{0,1,2\}$. As a consequence of isotropy, the two-point correlations can be written using two scalar functions $f(t,r)$ and $g(t,r)$, known as the longitudinal and transverse autocorrelation functions: $R_{00}/(u^\prime)^2 = f(r,t),\,\, R_{11}/(u^\prime)^2 = g(r,t),\,\, R_{22} = R_{11}, \mbox{ and }\,\, R_{ij} = 0 \,\mbox{ if }\, i \neq j$. 

\paragraph{Barycentric triangle} The Barycentric triangle is an anisotropy-invariant map, proposed by~\citet{BKDZ2007}. It defines a domain within which all realizable Reynolds stress invariants must lie. This representation overcomes the drawbacks of the maps proposed by Lumley and Newman, where the anisotropy invariants (II, III) are non-linear functions of the stresses. The barycentric map provides a non-distorted visual representation of anisotropy in turbulent quantities based on the convex combination of scalar metrics depending on eigenvalues. The Reynolds stress tensor $\bm{\tau}$ can be decomposed into isotropic and anisotropic parts as follows, 
\begin{equation}
    \bm{\tau} = 2 k \left( \dfrac{1}{3} \bm{I} + \bm{A} \right), 
\end{equation}
where $\bm{A}$ can be written as $\bm{V} \Lambda \bm{V}^{T}$, with $\bm{V}$ is a set of eigenvectors and $\Lambda$ the diagonal matrix composed of ordered eigenvalues ($\lambda_1 \geq \lambda_2 \geq \lambda_3$). In the Barycentric triangle, those eigenvalues are mapped to the Barycentric coordinates as, 
\begin{align*}
    C_1 &= \lambda_1 - \lambda_2 \\
    C_2 &= 2\left( \lambda_2 -\lambda_3 \right)\\
    C_3 &= 3 \lambda_3 + 1\,. 
\end{align*}
Because $\lambda_1+\lambda_2+\lambda_3=0$, we get that $C_1+C_2+C_3=1$. Setting the triangle in a Cartesian coordinate system $(\xi,\eta)$, every point inside the triangle is a convex combination of the three vertices that define the triangle $(\bm{\xi}_{1c}, \bm{\xi}_{2c}, \bm{\xi}_{3c})$, 
\begin{equation*}
    \bm{\xi} = C_1 \bm{\xi}_{1c} + C_2 \bm{\xi}_{2c} + C_3 \bm{\xi}_{3c}. 
\end{equation*}

\begin{figure}[h!t]
    \centering
    \includegraphics[width=.7\linewidth]{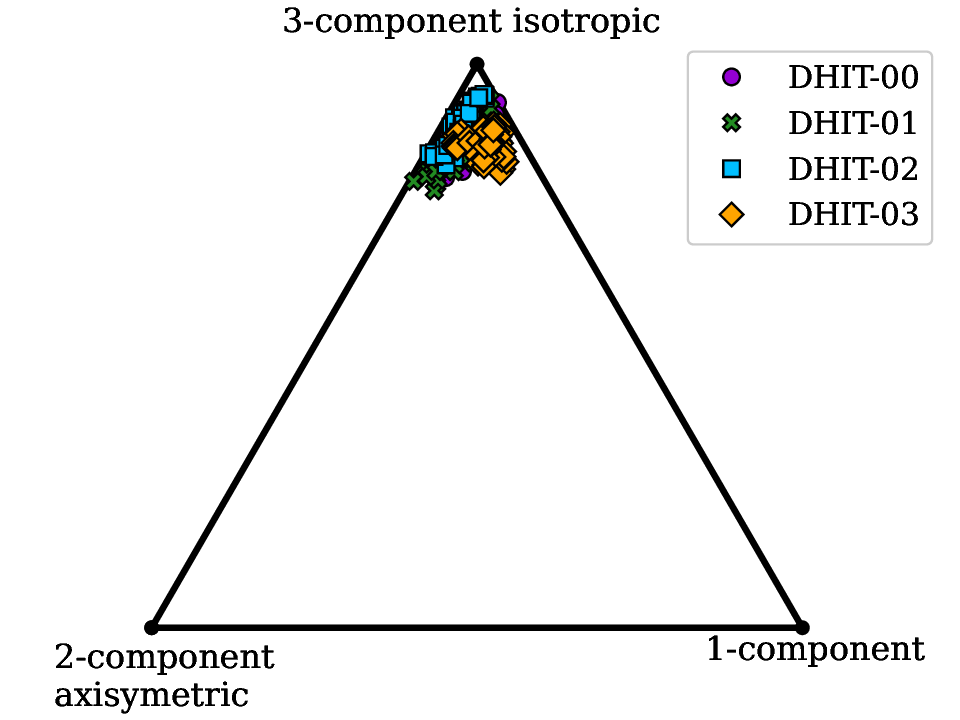}
    \caption{All physically realizable states of Reynolds stresses are enclosed in the barycentric triangle, with the three corners representing the limit states. The position within the triangle indicates the Reynolds stress anisotropy state.}
    \label{fig:barycentric}
\end{figure}

Figure~\ref{fig:barycentric} shows the Barycentric triangle for the four DHITs frozen at different time steps. All points are within the triangle and, as expected, close to the upper vertex, corresponding to 3-component isotropic turbulence. 

Therefore, all samples generated by the diffusion model that do not fall within the barycentric triangle are automatically discarded and considered to be unphysical.

\begin{figure}[h!t]
    \centering
    \includegraphics[width=.7\linewidth]{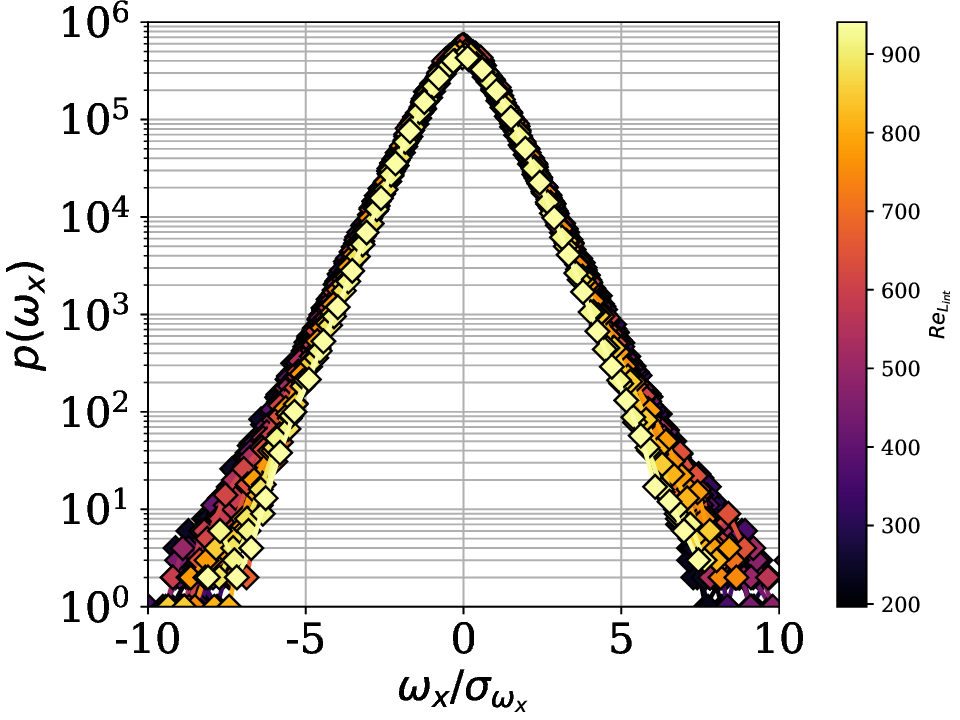}
    \caption{Distributions of the streamwise vorticity $\omega_x$ at various $Re_{L_{int}}$ for the database DHIT-03. }
    \label{fig:curl:distri}
\end{figure}

\paragraph{Vorticity probability distribution} The probability distribution of the vorticity $\bm{\omega} = \nabla \times \bm{u}$ is another important quantity to measure, since the formation of thin filaments creates large fluctuations in the local vorticity of the three-dimensional turbulence. Extreme events having a non-trivial tail distribution, the vorticity distribution has power-law tails, which should be reconstructed by the diffusion model. Figure~\ref{fig:curl:distri} shows the distribution of the streamwise vorticity at different $Re_{L_{int}}$ for the DHIT-03 database. As the Reynolds number decreases, the distribution widens while remaining symmetric around zero. Extreme events appear at the tails of the distribution for each Reynolds number.

\section{\label{sec:DDPM}Diffusion Models} 

Diffusion Models~\citep{SDWMG2015,HJA2020} (DM), have been developed to generate realistic data from a statistical distribution of interest $p(\bm{x})$. DM have recently demonstrated remarkable capabilities in image, video, or audio generation and have shown to be a notable prior for Bayesian inference. 

According to the formulation of~\citet{YSDKKEP2021}, a DM slowly corrupts samples $\bm{x}$ into random noise via a fixed forward diffusion (noising) process. This continuous-time forward process is a stochastic process equivalent to Brownian motion, governed by a Stochastic Differential Equation (SDE): 
\begin{equation}
    \mathrm{d}\bm{x}_t = \tilde{f}(t) \bm{x}_t \mathrm{d}t + \tilde{g}(t) \mathrm{d}\bm{w}_t, 
    \label{eq:forward}
\end{equation}
with $\bm{w}_t$ the standard Wiener process, $\bm{x}_t$ the perturbed samples at time $t\in [0,1]$, $\tilde{f}(t)\in \mathbb{R}$ and $\tilde{g}(t)\in \mathbb{R}^+$ the drift and the diffusion coefficients, respectively.

Following the notation of~\citet{KALA2022}, because the SDE is linear with respect to $\bm{x}_t$, the perturbation (or transition) kernel from $\bm{x}_0$ to $\bm{x}_t$ has the following general form, 
\begin{equation}
    p_{0t}(\bm{x}_t \vert \bm{x}_0) = \mathcal{N}\left(\bm{x}_t \,\vert\, s(t)\bm{x}_0, \,s^2(t)\sigma(t)^2\mathbf{I}\right),
    \label{eq:kernel}
\end{equation}
where $s(t)$ and $\sigma_t^2$ are the scaling and the scheduler. The forward diffusion process is chosen such that $\bm{x}_0 \sim p_{\mbox{data}}$, where $p_{\mbox{data}}$ denotes the data distribution, by imposing $s(0)=1$ and $\sigma(0) \ll 1$. The coefficients $\tilde{f}(t)$ and $\tilde{g}(t)$ are selected such that the initial samples $\bm{x}_0$ on the final perturbed samples $\bm{x}_1$ is negligible with respect to the noise level. Several noise schedulers satisfy these constraints but in this work, the standard Variance Preserving (VP) SDE is adopted. While alternative schedulers could have been considered (see Table~A~\ref{app:tab:SDE}), the primary objective here is to provide a proof-of-concept demonstration of applying diffusion-based methods to turbulence injection. Given the rapid progress in the field, a comprehensive comparison of different noise schedulers remains an open research direction and is left for future work.

Creating noise from data is relatively simple, but reconstructing realistic data from noise is a far more complex task. By starting with a sample from the prior distribution and reversing the diffusion process (backward diffusion), one can in principle obtain a sample from the data distribution $p_0$. The reverse process is also a diffusion process running backwards in time satisfying the following reversed SDE:
\begin{equation}
    \mathrm{d}\bm{x}_t = \left[\tilde{f}(t) \bm{x}_t + \dfrac{1+\eta^2}{2}\tilde{g}_t^2 \nabla_{\bm{x}_t} \log p(\bm{x}_t) \right]\mathrm{d}t + \eta \tilde{g}(t) \mathrm{d}\bm{w}_t,
\end{equation}
where $\eta \in \mathbb{R}^+$ controls the stochasticity. Solving the reverse SDE from $t=1$ to $0$ requires the access to the score function $\nabla_{\bm{x}_t} \log p(\bm{x}_t)$, which is unknown in practice, but can be approximated by a neural network via denoising score matching. Intuitively, the score function is a vector field, pointing towards higher data density at a given noise level.

This unknown score function is approximated by a deep neural network, denoted $s_\phi(\bm{x}_t,t)$ and usually referred to as a score network. Due to several instabilities in the training, the parameterization $\epsilon_\phi = -s(t)\sigma(t) s_\phi(\bm{x}_t,t)$ of the score network is considered and leads to a more stable and straightforward objective~\citep{RL2023},
\begin{equation}
    \mbox{arg}\,\min_\phi \mathbb{E}_{p(\bm{x})p(t)p(\epsilon)} \left[ \Vert \epsilon_\phi \left(s(t)(\bm{x}+\sigma(t)\epsilon), t \right) - \epsilon \Vert^2\right],
    \label{eq:loss}
\end{equation}
where $p(\epsilon)=\mathcal{N}(0,\mathbf{I})$. The above parametrization and its corresponding objective is not the only one proposed in the literature~\citep{HJA2020,SE2019,YSDKKEP2021,SME2021,KALA2022,LCBHNL2023,RALL2024} to accomplish this task. 

The performance of DMs comes with a major limitation: their slow sampling procedure, which typically requires hundreds to thousands of time discretization steps (i.e., steps required to solve the reverse SDE) of the learned diffusion process to achieve the desired accuracy. In~\citet{YSDKKEP2021}, black-box ODE solvers were used to speed up the sampling process by solving a marginal equivalent ODE known as Probability Flow (PF). In the present work, the predictor/corrector method presented by~\citet{RL2023} is considered. This method corresponds to the exponential integrator discretization scheme proposed by~\citet{ZC2023}, combined with several Langevin Monte Carlo steps to avoid error accumulation. The DM trained in this work follows~\citet{RL2023,RALL2024} as implemented in their GitHub repository~\citep{sda,Azula}.

\subsection{Guidance}
In the database, there are multiple DHITs at different TKE levels. Consequently, the objective is to sample a DHIT at a given level of TKE. This task can be accomplished by conditioning the diffusion model. According to~\citet{HCL2024}, two major approaches are available: \textit{classifier guidance} and \textit{classifier-free guidance}. 

Classifier guidance has been proposed by~\citet{DN2024} to enhance the sample quality of a diffusion model by using an dedicated classifier. \citet{HS2022} showed that a purely generative model can provide guidance without such a classifier. In their work, a single diffusion model is trained with conditioning dropout. Therefore, some percentage of the time ($10$-$20\%$), the class conditioning $c$ is set to a '\textit{none}' value (i.e., absence of conditioning). The diffusion model parametrizes simultaneously a conditional and an unconditional model. Their resulting scores, i.e., $\nabla_{\bm{x}_t} \log p(\bm{x}_t)$ and $\nabla_{\bm{x}_t} \log p(\bm{x}_t,c)$, respectively, are combined to achieve a trade-off between diversity and quality similar to that obtained with classifier guidance. No additional cost is associated with training a classifier model. 

\begin{figure}[h!t]
    \centering
    \includegraphics[width=.7\linewidth]{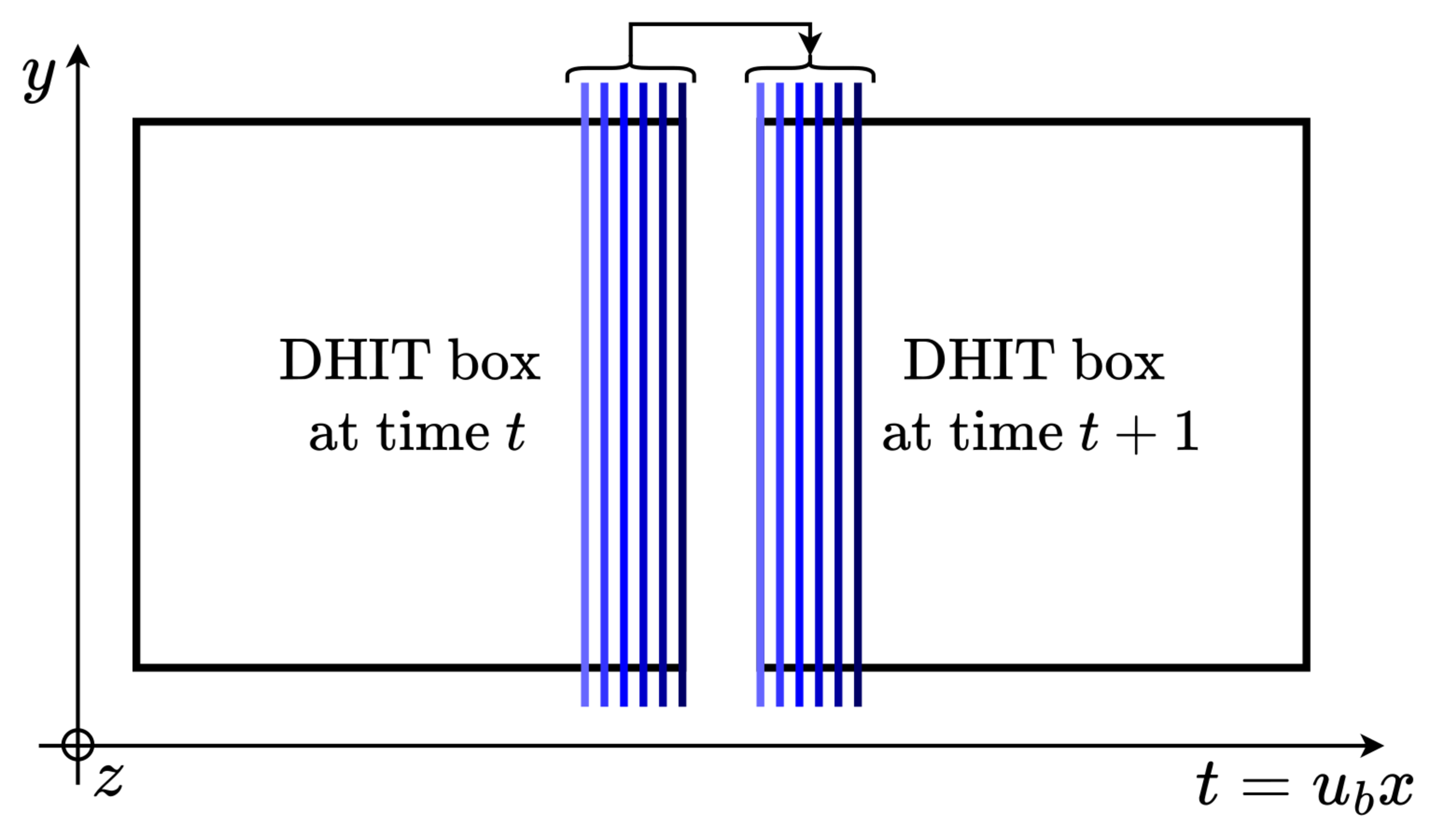}
    \caption{Schematic of the mask used in the MMPS method to impose continuity between successive boxes.}
    \label{fig:MMPS}
\end{figure}

For sampling, a barycentric combination of the conditional and the unconditional score function is used. By taking the derivative with respect to $\bm{x}_t$ of the logarithm of the Bayes' rule of $p(c\vert \bm{x}_t)$, one gets, 
\begin{equation*}
    \nabla_{\bm{x}_t} \log p(c\vert \bm{x}_t) = \nabla_{\bm{x}_t} \log p(\bm{x}_t \vert c) - \nabla_{\bm{x}_t} \log p(\bm{x}_t).
\end{equation*}
This expression is substitute into the formula of the classifier guidance, 
\begin{align*}
    \nabla_{\bm{x}_t} \log p_\gamma(\bm{x}_t \vert c) & = (1-\gamma) \nabla_{\bm{x}_t} \log p(\bm{x}_t) \\
    & + \gamma \nabla_{\bm{x}_t} \log p(\bm{x}_t \vert c)
\end{align*}
where $\gamma$ is the guidance scale. The classifier-free guidance method is simpler to implement and use. It requires only one line of change during training to randomly drop the class condition $c$, and during sampling by mixing the conditional and unconditional score estimates. For these reasons,  classifier-free guidance is considered in this work.

\subsection{Moment Matching Posterior Sampling}

In the precursor procedure, to avoid re-injecting the same box over and over, \citet{RTBH2023} proposed to apply a series of translations and rotations to the DHIT box to generate new realizations that are then concatenated together using the blending procedure proposed by~\citet{XNL2012}. Generative models are interesting because they can generate new samples, thus eliminating the need for translation and rotation. Nevertheless, the continuity between consecutive boxes should be controlled. The first approach to ensure the continuity is to reuse the blending procedure of~\citet{XNL2012}, but it requires a step to project the synthetic velocity field into zero-divergence space. The second approach uses the Moment Matching Posteriori Sampling (MMPS)~\citep{RL2023,RALL2024}. This method allows to sample from the posterior distribution without the need to retrain the diffusion model. The objective is to generate a new DHIT box knowing the $m$ last $(y{-}z)$ slices of the previous boxes (see Figure~\ref{fig:MMPS}). The method requires the definition of a measurement function $\mathcal{A}$ and a Gaussian forward process $p(x\vert y) = \mathcal{N}(y \vert \mathcal{A}(\bm{x}), \Sigma_y)$. For the present problem, the measurement operator is a mask that cancels all but the first $m$ slices. The operator is a sparse matrix $\bm{A} \in \mathbb{R}^{(3n^3)\times (3n^3)}$, where $n=64$ is the box resolution and contains only $3mn^2$ non-zero elements. % differentiable !!!! 

The application of the MMPS procedure necessitates the definition of the number of slices $m$ and the variance $\Sigma_y$. In Section~\ref{sec:res}, $m$ and $\Sigma_y$ are set to $16$ and $10^{-4}$, respectively. These values are obtained by grid search and are those that best preserve the statistics of interest defined in Section~\ref{sec:data}.

% ---
% Training on 64³ and on multiple Re_l 
% A priori results (spectrum and correlation functions)
% A posteriori results after injection (anisotopy, spectrum, integral length scales, ...)
% ---
\section{\label{sec:res}Diffusion models applied to multiple DHIT boxes}

In this section, the 214 DHIT boxes, each labeled with a unique pair $(\mbox{TKE}, L_{int})$, are used to train a CF-DM, conditioned by the Reynolds number defined as, 
\begin{equation}
    Re_{L_{int}} = \sqrt{\dfrac{2}{3}k}\dfrac{L_{int}}{\nu},
\end{equation}
where $\nu$ is the kinematic viscosity. Although the integral length scale is an important quantity that determines the size of the vortices in turbulent fields, the range of $L_{int}$ is not sufficient to condition the model with it. Therefore, both the TKE level and the integral length scale are embedded in this Reynolds number. 

The database cannot be used directly without analyzing the distribution of the Reynolds number $Re_{L_{int}}$ associated with each box. A good practice is to have a uniform data distribution to ensure an equivalent representation of each box during training. Figure~\ref{fig:histo} shows the distribution of the database in green. The distribution of $Re_{L_{int}}$ is not uniform. Training a model on this data will lead to a good prediction for $200\leq Re_{L_{int}} \leq 450$, but would have difficulty predicting $Re_{L_{int}}\geq 800$. Careful selection has resulted in the blue histogram showing a more uniform distribution. A similar technique is applied to the validation data. A total of 62 boxes and 17 boxes are used for training and validation.

\subsection{\label{subsec:apriori}\textit{A priori} validation of the CF-DM}

Table~\ref{tab:CFDM} summarizes the hyperparameters chosen to train the CF-DM. The model is also trained in parallel on 4 GPUs (Nvidia A100), yielding a restitution time of 27 minutes to complete one epoch.

\begin{minipage}{\textwidth}
    \begin{minipage}[b]{0.49\textwidth}
    \centering
    \begin{tabular}{c|c}
        \hline \hline 
        \textbf{Hyper-parameters} & \textbf{Values} \\
        \hline 
        \textbf{Training size} & $136{,}102$ \\
        \textbf{Validation size} & $9{,}329$ \\
        \textbf{Batch size} & $7$ \\
        \textbf{Input size} & $3 \times 64 \times 64 \times 64$ \\
        \textbf{Output size} & $3 \times 64 \times 64 \times 64$ \\
        \textbf{$\#$ translations} & $14$ \\
        \hline 
        \textbf{SDE} & VPSDE (Table~A~\ref{app:tab:SDE}) \\
        \textbf{Scheduler} & Cosine \\
        \hline 
        \textbf{Architecture} & UNet~\citep{RALL2024} \\
        \textbf{Number of parameters} & $8{,}030{,}339$ \\
        \hline 
        \textbf{Learning rate} & $10^{-4}$\\
        \textbf{Optimizer} & \texttt{torch.optim.AdamW} \\ 
        \hline \hline
    \end{tabular}
    \captionof{table}{\label{tab:CFDM}Hyper-parameters employed to train the CF-DM on  DHIT boxes at multiple $Re_{L_{int}} \in [196.58, 1302.16]$.}
\end{minipage}
\hfill
\begin{minipage}[b]{0.49\textwidth}
    \centering
    \centering
    \includegraphics[width=\linewidth]{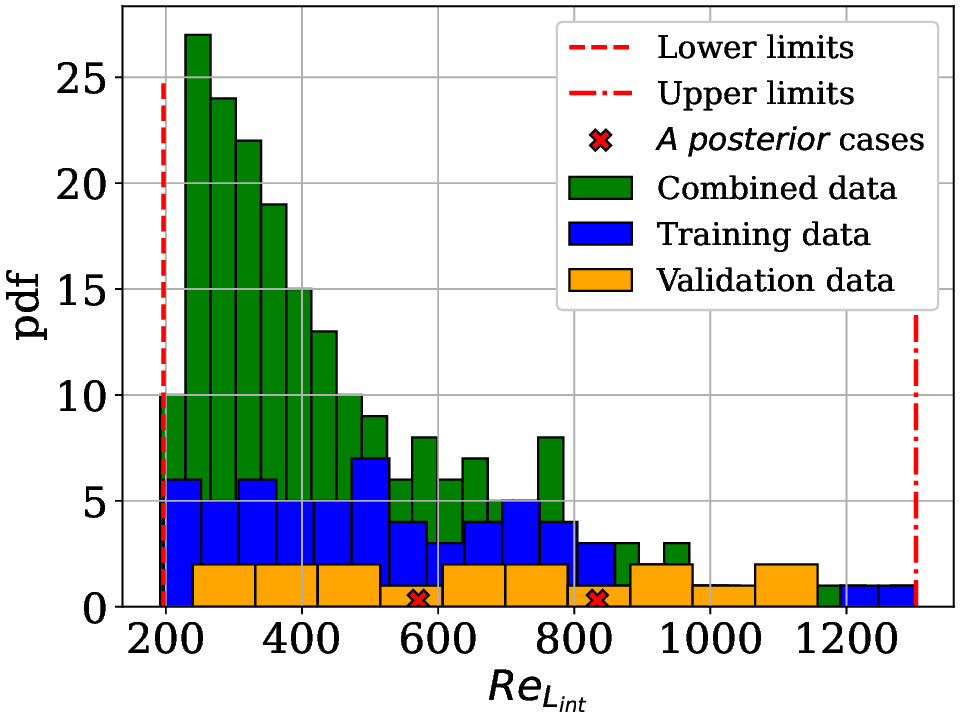}
    \captionof{figure}{\label{fig:histo}Distribution of the Reynolds number $Re_{L_{int}}$ for the $214$ DHITs, the training and validation data in green, blue and orange.}
    \end{minipage}
\end{minipage}
  
The frozen model is used to generate new samples using the predictor/corrector method proposed by~\citet{RL2023}. This method contains two hyper-parameters: (1) the number of Langevin Monte Carlo (LMC) corrections $\lambda$ and (2) the discretized reverse SDE step $\tau$. These parameters can affect the quality of the generated samples. According to~\citet{RL2023}, in the limit of an infinite number of LMC corrections combined with a sufficiently small discretized step, the samples are guaranteed to follow the distribution implicitly defined by the score approximation at each time $t$. Due to the conditioning, this sampling method contains an additional parameter: the guidance factor $\gamma$. The optimal $\lambda^\ast$, $\tau^\ast$, and $\gamma^\ast$ are those how minimize the following loss, 
\begin{equation}
    \mathcal{L}(\lambda,\tau,\gamma) = \alpha_{k} h_{k} + \alpha_{E_\kappa} h_{E_\kappa} + \alpha_{f,g} h_{f,g} + \alpha_{\triangle} h_{\triangle},
    \label{eq:hppc}
\end{equation}
where $h_{k}$, $h_{E_\kappa}$, $h_{f,g}$, and $h_{\triangle}$ are defined as, 
\begin{align*}
    h_{k} &= \dfrac{\vert \hat{k}-k\vert}{k}, \, h_{\triangle} = \sum_i 1_{(\xi_i,\eta_i) \notin \triangle}\\
    h_{E_\kappa} &= \dfrac{\frac{1}{n/2}\sum_i (\hat{E}(\kappa_i)-E(\kappa_i))}{\frac{1}{n}\sum_i E(\kappa_i)}\\
    h_{f,g} &= \dfrac{1}{n} \sum_i \left( \left( \hat{f}(r_i) - f(r_i) \right)^2 + \left( \hat{g}(r_i) - g(r_i) \right)^2 \right) 
\end{align*}
and the coefficients $\alpha_{k}$, $\alpha_{E_\kappa}$, $\alpha_{f,g}$, and $\alpha_{\triangle}$ are set to $1$, $1$, $100$, and $1/n_s$, with $n_s$ is the number of generated samples. A grid search is performed over $[\lambda] \times [\tau] \times [\gamma] = [1,2,4,6,8] \times [0.01,0.1,0.25,0.5,0.75,1] \times [1,2,3,4,5]$. The optimal hyperparameters are $\lambda^\ast=4$, $\tau^\ast=0.1$, and $\gamma^\ast=2$.

The model is \textit{a priori} validated on two Reynolds numbers $Re_{L_{int}}$ that are within the training limits, i.e., $196.58\leq Re_{L_{int}} \leq 1302.16$. The model is pushed further by analyzing the quality of the results with the conditioning $Re_{L_{int}}$ taken outside the training limits, i.e., $Re_{L_{int}}<196.58$ or $Re_{L_{int}}>1302.16$. 

\paragraph{Validation inside and outside the training limits} The CF-DM is first validated on two Reynolds numbers $Re_{L_{int}}=\{834,\,571\}$ chosen within the training limits. The same model is also validated on two Reynolds numbers $Re_{L_{int}}=\{1374,\,192\}$ chosen outside the training limits. This is not considered aggressive extrapolation, but it already evaluates how well the model performs at the training limits.

The generated fluctuations are analyzed using the four statistics of interest, described in Section~\ref{sec:data}. 

\begin{figure}[h!t]
    \centering
    \includegraphics[width=0.7\linewidth]{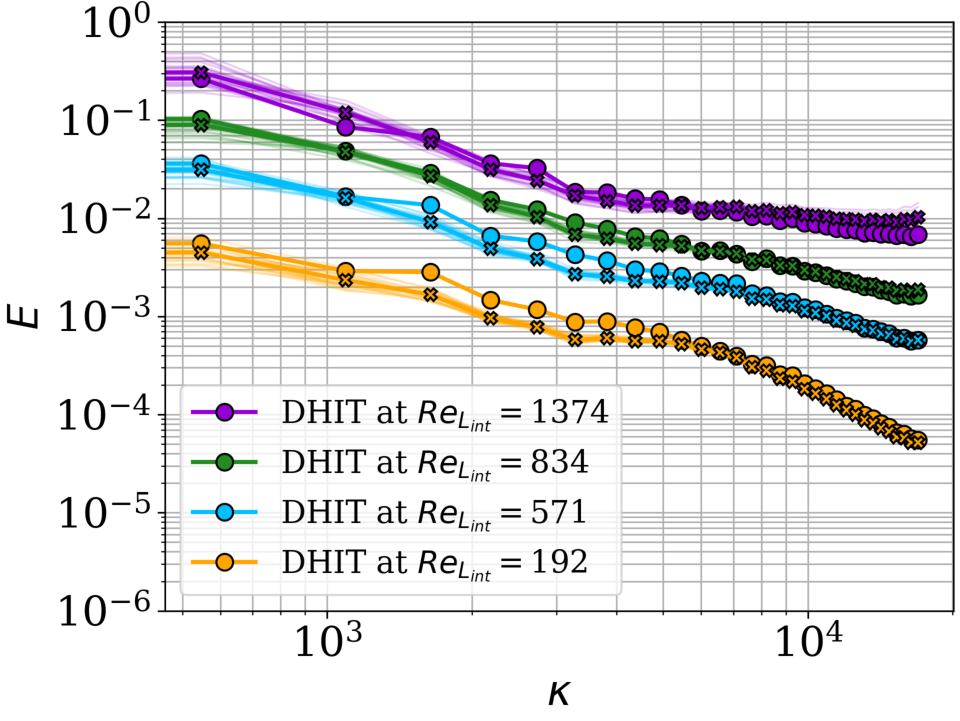}
    \caption{\textbf{\textit{A priori} assessment of the CF-DM conditioned at $\bm{Re_{L_{int}}=\{1374,834,571,192\}}$.} The energy spectrum at four TKE levels with the ground truth marked by solid circles and the generated samples marked by solid crosses of the same color. }
    \label{fig:multi:priori:inside:spectrum}
\end{figure}

Figure~\ref{fig:multi:priori:inside:spectrum} shows the energy spectrum in plain circles for the ground truth, and in plain crosses for the conditionally generated samples: the green and blue curves correspond to $Re_{L_{int}}=\{834,\,571\}$, while the purple and orange curves correspond to $Re_{L_{int}}=\{1374,\,192\}$, respectively. Within the training limits, a good agreement between the two spectra is observed, with a small underestimation of the inertial range. Similar observations are made outside the training limits. A total of 24 boxes are generated per $Re_{L_{int}}$, and such a diversity is observed at the largest scales on each spectrum.

\begin{figure}[h!t]
    \centering
    \includegraphics[width=0.7\linewidth]{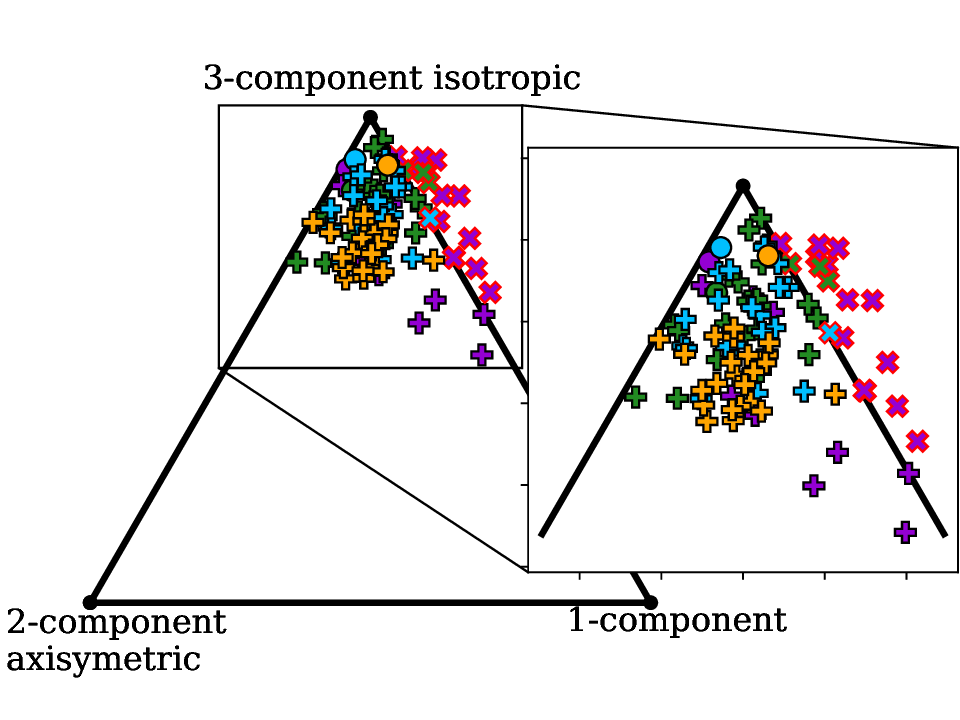}
    \caption{\textbf{\textit{A priori} assessment of the CF-DM conditioned at $\bm{Re_{L_{int}}=\{1374,834,571,192\}}$.} Barycentric triangle with the circles indicating the ground truth and the crosses indicating the samples obtained from the CF-DM. Color mapping follows Figure~\ref{fig:multi:priori:inside:spectrum}.}
    \label{fig:multi:priori:inside:bary}
\end{figure}

At the highest $Re_{L_{int}}$, several boxes are outside the Barycentric triangle and should be considered as unrealizable turbulence (see Figure~\ref{fig:multi:priori:inside:bary}). For the other three Reynolds numbers, the points are clustered near the upper corner, indicating isotropic Reynolds stress states. Conversely to the training on a single box, a largest variance is observed in the Barycentric map. The points outside the triangle are simply discarded as they are considered unphysical realization states of the Reynolds stress. The possibility of guiding the diffusion model to generate turbulence within the Barycentric map is discussed in the Perspectives section. \hlwhite{Discuss more the limitation of a pure data-driven model in regards of this plot and the possible need to physically constraint the DM.}

\begin{figure*}[h!t]
    \centering
    \includegraphics[width=.99\linewidth]{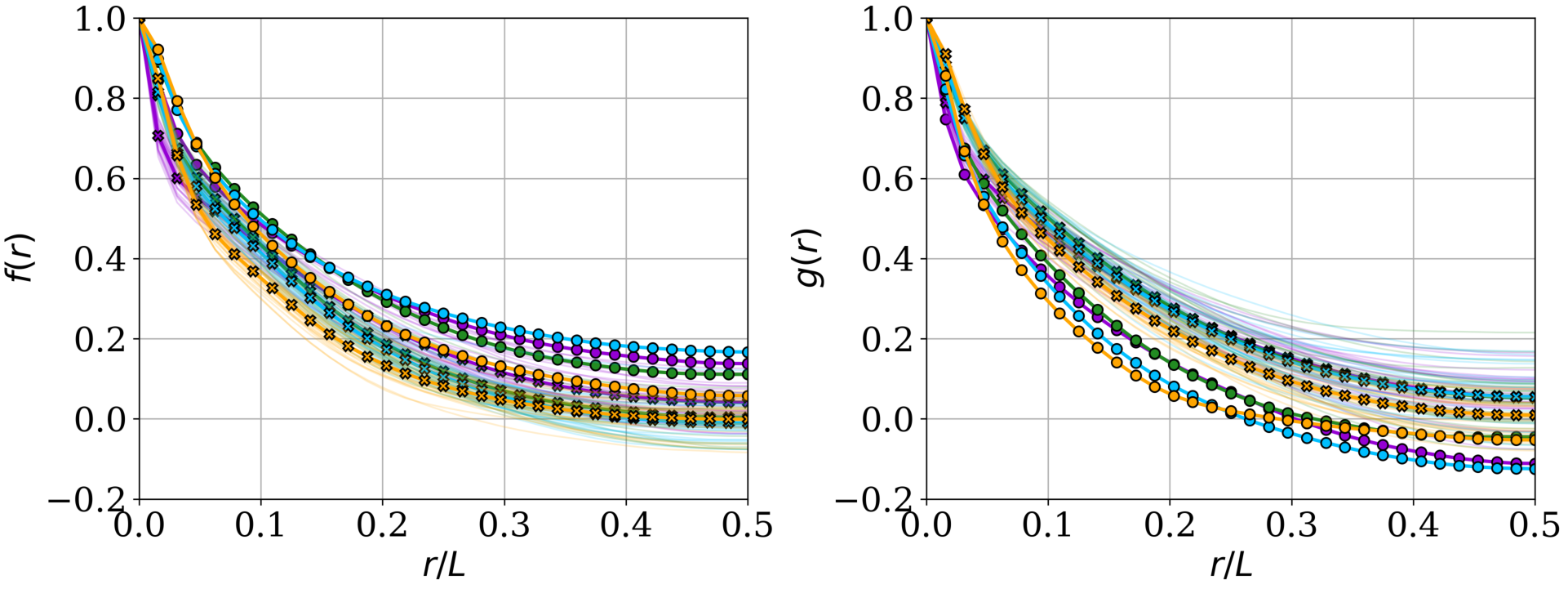}
    \caption{\textbf{\textit{A priori} assessment of the CF-DM conditioned at $\bm{Re_{L_{int}}=\{1374,834,571,192\}}$.} The longitudinal and transverse correlation functions, on the left and right, respectively, with the ground truth marked in plain circles and the generated samples in plain crosses. The color mapping is similar to Figure~\ref{fig:multi:priori:inside:spectrum}.}
    \label{fig:multi:priori:inside:corr}
\end{figure*}

Figure~\ref{fig:multi:priori:inside:corr} shows the longitudinal and transverse correlation functions. In all CF-DM trainings, we observed an underestimation of $f(r)$ and an overestimation of $g(r)$.
%Of all the trained CF-DM, the correlation functions were the hardest to fit, and an underestimation of $f(r)$ and an overestimation of $g(r)$ were always observed. Note also the wide variability of the profiles. The network has some difficulties to distinguish between the four Reynolds numbers, which are mainly characterized by a different level of TKE, but also by a small variation of the integral length scales. 

\begin{figure*}[h!t]
    \centering
    \includegraphics[width=.99\linewidth]{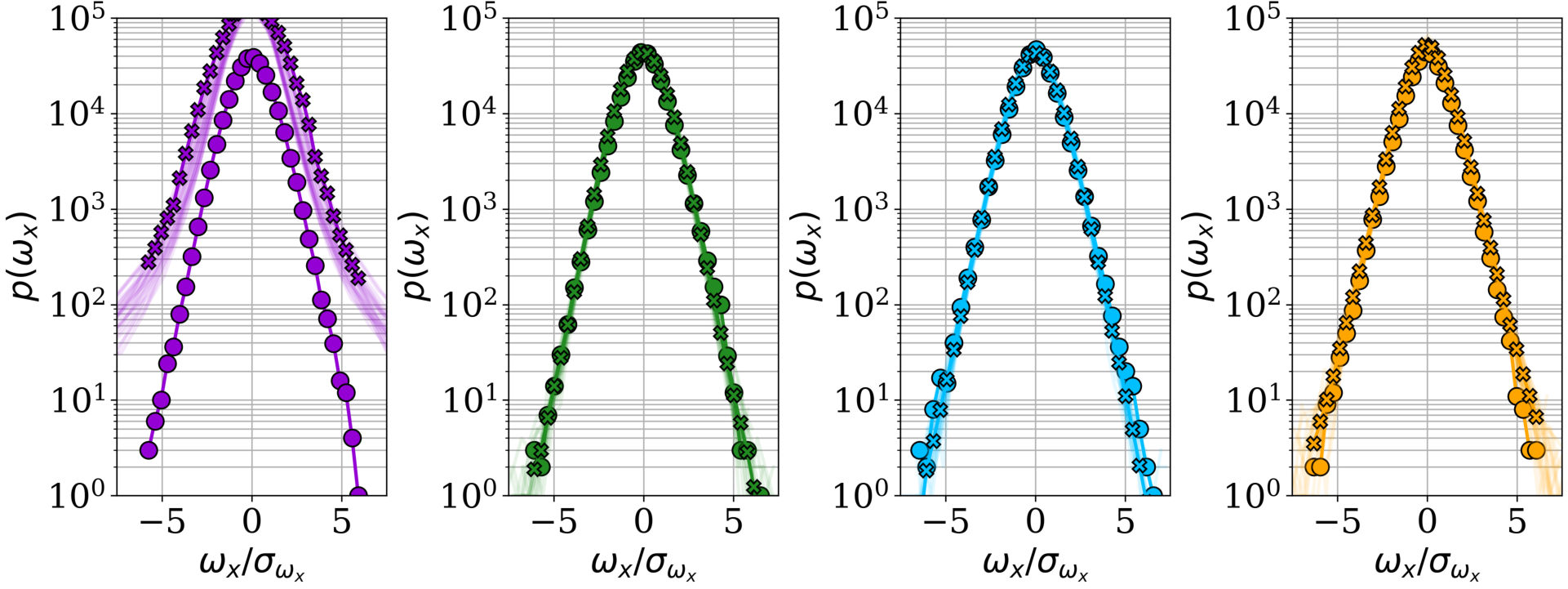}
    \caption{\textbf{\textit{A priori} assessment of the CF-DM conditioned at $\bm{Re_{L_{int}}=\{1374,834,571,192\}}$.} The distribution of the streamwise vorticity $\omega_x$. The color mapping is similar to Figure~\ref{fig:multi:priori:inside:spectrum}.}
    \label{fig:multi:priori:inside:curl}
\end{figure*}

Figure~\ref{fig:multi:priori:inside:curl} shows the distribution of the streamwise vorticity, where a good agreement with the ground truth is observed for $Re_{L_{int}}=\{834,571,192\}$, even in the tails of the distribution. At the highest Reynolds number, however, the vorticity distribution is strongly overestimated.

The model is able to generate decent synthetic turbulence at different $Re_{L_{int}}$ located within the training limits. \hlwhite{Conditioning the CF-DM with a higher Reynolds number than the training limits produces unphysical fluctuations according to the Barycentric map and the vorticity distribution.} However, conditioning the CF-DM with a lower Reynolds number than the training limits produces samples that are coherent. It seems easier for the model to dissipate small structures than to reconstruct them, since the DHIT at lower $Re_{L_{int}}$ can be seen as a filtered version of the DHIT at higher $Re_{L_{int}}$.

The most difficult statistical quantity to obtain from the samples is the autocorrelation function. The model seems to have difficulty distinguishing between the integral length scales and predicts the same correlations at each $Re_{L_{int}}$. Otherwise, the other three target statistics are well reproduced and good agreement with the ground truth is observed.

\subsection{\label{subsec:multi:post}\textit{A posteriori} assessment of the CF-DM}

\myhl{The \textit{a posteriori} assessment requires the injection of the generated turbulence into a free domain using the inlet boundary condition defined in Section~\ref{subsec:solver}. The computational domain corresponds to a long parallelepiped with the same pitch and span length as the DHIT box, denoted $L$ (see Figure~\ref{fig:framework}). The streamwise length ($L_x$) is equal to $5.88\,L$. The resolution in the free domain is similar to that of the DHIT, except near the outlet where the mesh cells are enlarged to damp the fluctuations. The mesh contains a total of $156{,}585$ hexahedra and is subdivided into 256 partitions. The simulation is performed with the solver Argo-DG at a polynomial order $p=4$ for a total of about $20,000,000$ degrees of freedom. The time step is set to $10^{-6}$ [s] to ensure a maximum CFL number of about $0.68$ and thus the correct resolution of turbulent structures. }

\myhl{The inlet boundary condition (Section~\ref{subsec:solver}) imposes a total pressure of $182{,}800$ [Pa] and a total temperature of $413.3$ [K]. At the outlet, a static pressure of $179{,}996$ [Pa] is imposed. The other four boundaries (i.e., the boundaries tangential to the streamwise direction) are treated as periodic. A sponge layer is applied at the outlet to avoid spurious reflections of the solution back into the main domain. }

The physical representativeness of the generated samples, conditioned on the Reynolds number within the training limits, is evaluated \textit{a posteriori} by injecting the fluctuations into the free domain. The difference is the level of turbulence intensity injected at the domain inlet. For $Re_{L_{int}}=834$ (see Section~\ref{subsubsec:multi:post:850}), the turbulence intensity is about $23.3\%$, while for $Re_{L_{int}}=571$ (see Section~\ref{subsubsec:multi:post:1500}), the turbulence intensity is $14.7\%$. In both cases the statistics are accumulated over $10\,t_c$.

\myhl{The capability of DM to generate new boxes from the same approximated distribution is used to avoid the preprocessing step translating and rotating the unique DHIT box to reduce the time correlation while injecting the fluctuations. Two methods to concatenate these generated boxes are tested and compared against the original precursor method. The first method is the one that simply blend consecutive boxes using two blending functions that conserves the turbulent kinetic energy. The second method uses the MMPS presented in Section~\ref{sec:DDPM} with $m=16$ and $\Sigma_y=10^{-4}$. }

\begin{figure}[h!t]
    \centering
    \includegraphics[width=0.6\linewidth]{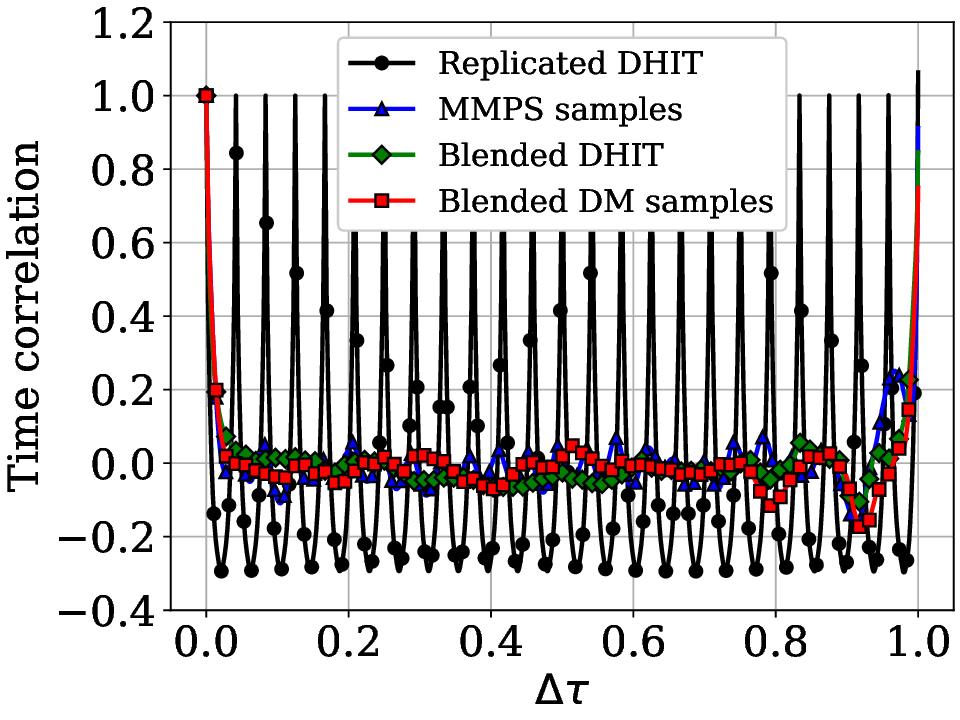}
    \caption{Time correlation computed for different blending techniques, the samples are obtained by conditioning the CF-DM at $Re_{L_{int}}=571$ using the following hyper-parameters: $m=16$, $\Sigma_y=10^{-4}$, $\lambda^\ast=4$, $\tau^\ast=0.1$, and $\gamma^\ast=2$.}
    \label{fig:multi:post:time}
\end{figure}

Figure~\ref{fig:multi:post:time} shows the time correlation computed over the 24 generated boxes. The simplest approach is to replicate the DHIT box (black circle curve), but it produces large correlation peaks. The method proposed by~\citet{RTBH2023} \myhl{weakens} these peaks by translating, rotating, and blending the original DHIT box (plain diamond green curve). Only the blending procedure is applied to 24 synthetic DHIT boxes sampled from the CF-DM. Indeed, a DM is a stochastic model that can generate multiple boxes with similar statistics but different instantaneous snapshots, while with the original precursor method, only one box at a given TKE level is available per simulation. By blending the 24 boxes, the resulting signal shows no correlation peak (plain red square curve). The MMPS method produces decorrelated boxes since the correlation fluctuations remain below $10\%$. The MMPS method can also be used to impose the periodicity between the last and the first boxes to avoid brutal switching when reading the box during the simulation. The MMPS is particularly interesting when the model is implemented directly in the flow solver, since the model can generate boxes over and over again at infinity reducing the storage need. 

\subsubsection[Freestream turbulence injection at TI $=23.3\%$]{Freestream turbulence injection at TI $\boldsymbol{=23.3\%}$} 
\label{subsubsec:multi:post:850}

The turbulence injection is first performed using samples generated at $Re_{L_{int}}=834$, corresponding to a turbulence intensity of $23.3\%$ at the inlet plan of the main computational domain. The objective is to evaluate the streamwise evolution of four quantities: the TKE, the integral length scale, the two-dimensional energy spectrum, and the diagonal terms of the Reynolds stress tensor (i.e., the anisotropy). 

\begin{figure}[h!t]
    \centering
    \includegraphics[width=0.7\columnwidth]{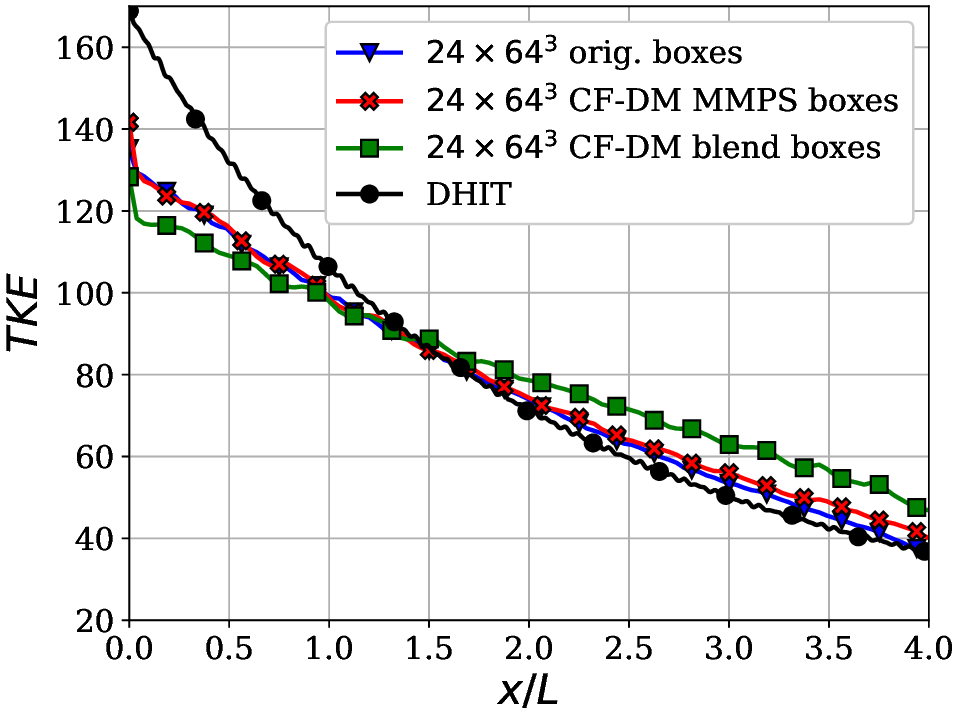}
    \caption{\label{fig:multi:850:post:TKE}\textbf{Freestream turbulence injection at TI$\bm{=23.3\%}$ to \textit{a posteriori} assess the CF-DM conditioned with $\bm{Re_{L_{int}}=834}$.} Streamwise evolution of the TKE.}
\end{figure}

Figure~\ref{fig:multi:850:post:TKE} shows the streamwise evolution of the TKE. The black circle curve corresponds to the TKE of the DHIT simulation, where time is converted to space evolution using the mean inlet velocity. The blue triangle curve corresponds to the original precursor method, where a significant drop in the inlet plane is observed. The abrupt energy loss can have several reasons: 1) could be an interpolation error between the probe grid, on which the velocity fluctuations are stored, and the degrees of freedom at the inlet plane, 2) could be a TKE-unconservative boundary condition resulting in a lag in NS resolution at the inlet, 3) could be a violation of the Taylor hypothesis at higher TI. 

In Figure~\ref{fig:multi:850:post:TKE}, the boxes blended using the MMPS (in red cross-shaped curve) produce a TKE comparable to the original precursor method. A similar decrease in TKE is also observed at the inlet. For the boxes concatenated using the blending functions (in green square-shaped curve), a large drop is observed at the inlet. Even if the blending procedure tends to preserve the TKE level, its original level of $158.1$ drops to $155.9$ after blending the boxes. The original precursor method has a TKE level of $161.9$ after blending, therefore, even before injecting the boxes into the domain, there is already a $3.7\%$ difference between the TKE levels. This gap is corrected after one box length but yields an over estimated TKE after two box lengths.

\begin{figure}[h!t]
    \centering
    \includegraphics[width=0.7\columnwidth]{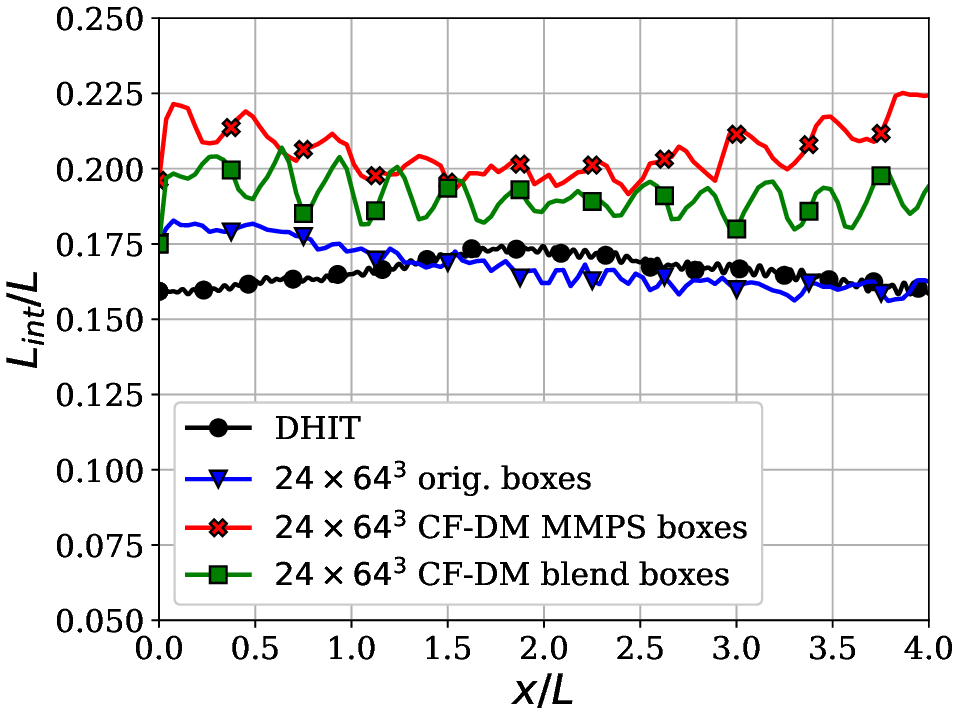}
    \caption{\label{fig:multi:850:post:ILgth}\textbf{Freestream turbulence injection at TI$\bm{=23.3\%}$ to \textit{a posteriori} assess the CF-DM conditioned with $\bm{Re_{L_{int}}=834}$.} Streamwise evolution of the longitudinal length scale.}
\end{figure}

Figure~\ref{fig:multi:850:post:ILgth} shows the evolution of the integral length scale $L_{int}$ as defined in Equation~\ref{eq:lint}. As already observed in the \textit{a priori} assessment, the CF-DM does not correctly predict the autocorrelation function, and the integral length scale is overestimated (by about $14\%$). For both the MMPS generated and the blended boxes, the integral length scale remains approximately constant after about $20\%$ of the box length. 

\begin{figure*}[h!t]
    \centering
    \includegraphics[width=\linewidth, trim=0 220 210 0, clip=true]{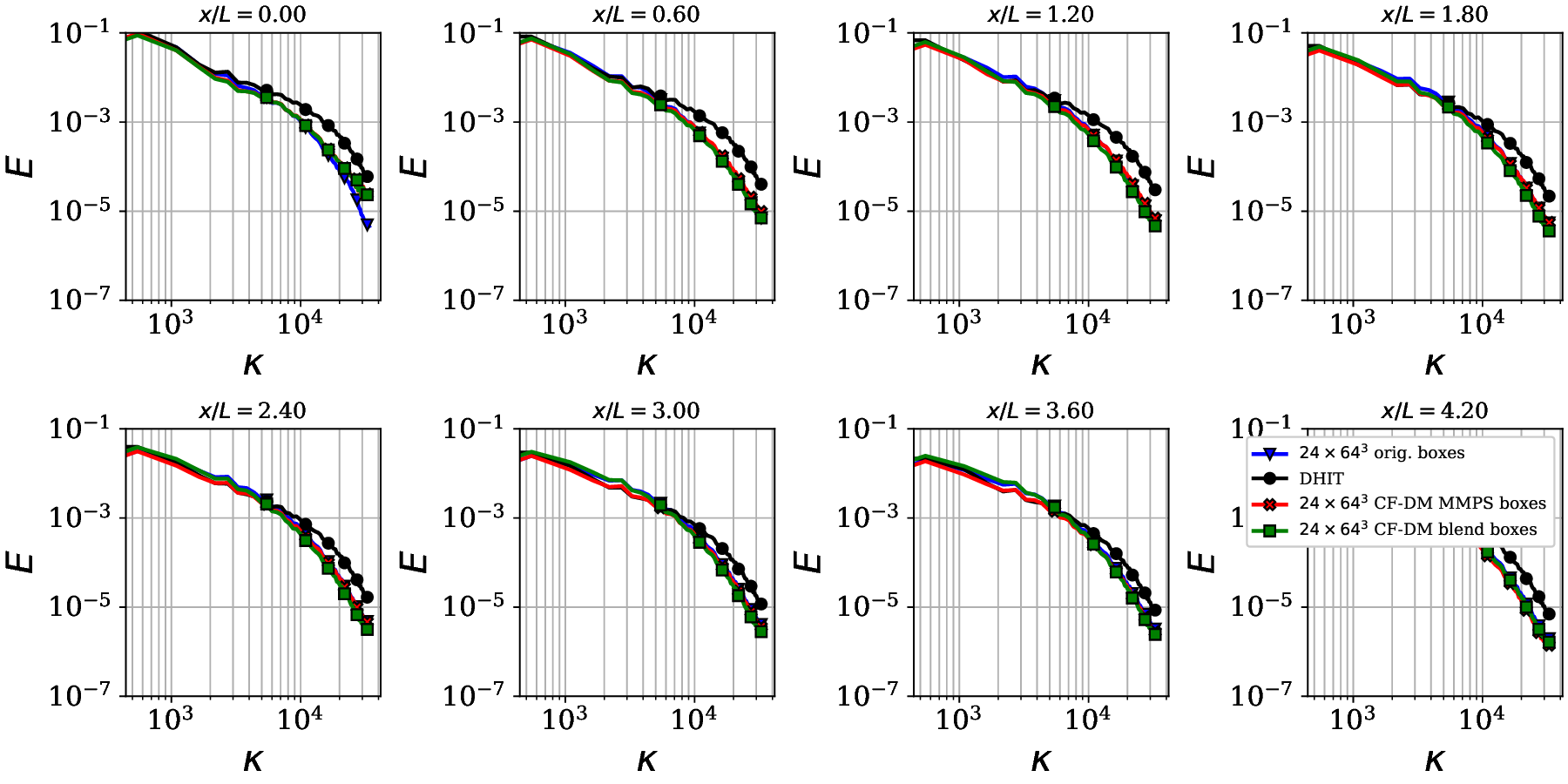}
    \caption{\label{fig:multi:850:post:spectrum}\textbf{Freestream turbulence injection at TI$\bm{=23.3\%}$ to \textit{a posteriori} assess the CF-DM conditioned with $\bm{Re_{L_{int}}=834}$.} 2D energy spectrum at three streamwise locations $x/L=\{0,0.6,1.2\}$ in the free domain, the curves colors following Figure~\ref{fig:multi:850:post:TKE}.}
\end{figure*}

Figure~\ref{fig:multi:850:post:spectrum} shows the evolution of the two-dimensional energy spectrum at three streamwise locations. In less than one box length, the energy spectrum of the turbulence injected by the true DHIT (original precursor method) and the CF-DM (MMPS and blended) generated boxes overlap. \myhl{None of the energy spectra of the injected fluctuations matches the DHIT 2D energy spectrum, because for $x/L<1.5$ the TKE retrieved in the free-domain is smaller than the expected TKE level.}

\begin{figure}[h!t]
    \centering
    \includegraphics[width=0.7\columnwidth]{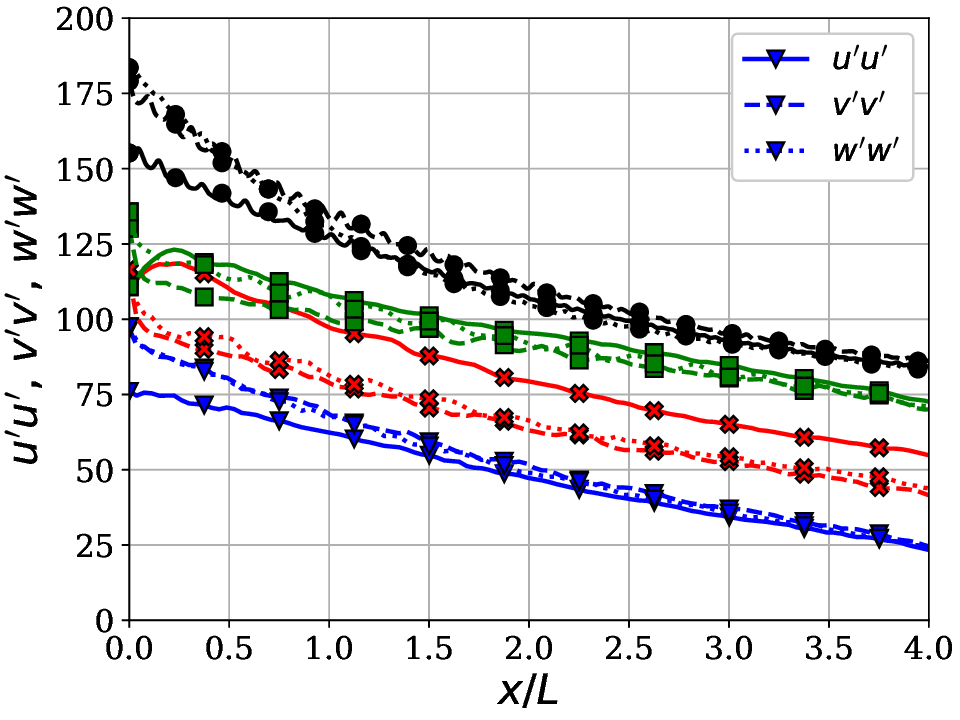}
    \caption{\label{fig:multi:850:post:anisotropy}\textbf{Freestream turbulence injection at TI$\bm{=23.3\%}$ to \textit{a posteriori} assess the CF-DM conditioned with $\bm{Re_{L_{int}}=834}$.} Streamwise evolution of the Reynolds stress components in the free domain, the different sets of curves are shifted by $10$ for readability, the color mapping is similar to Figure~\ref{fig:multi:850:post:TKE}.}
\end{figure}

Figure~\ref{fig:multi:850:post:anisotropy} shows the diagonal terms of the Reynolds stress tensor. Both the original precursor method and DHIT exhibit anisotropy at the inlet, which quickly disappears after a box length. The $u^\prime u^\prime$ component appears smaller than the other two. \hlwhite{The anisotropy is clearly visible for the MMPS generated boxes, but in this case the $u^\prime u^\prime$ component is dominant. The imposition of two-dimensional $(y-z)$ snapshots possibly induces unwanted forcing in the streamwise direction.} \myhl{A potential solution would be to add physical constraints to the generation process, as discussed in Perspectives}. For the blended boxes, no anisotropy is observed except near the inlet plane. Appendix~\ref{subsec:app:post:res} elaborates on these aspects, with Figure~A~\ref{app:fig:multi:850:post:bary} highlighting the anisotropy states of the Reynolds stress tensor inside the Barycentric triangle.

\subsubsection[Freestream turbulence injection at TI $=14.7\%$]{Freestream turbulence injection at TI $\boldsymbol{=14.7\%}$} 
\label{subsubsec:multi:post:1500}

The turbulence injection is performed using samples generated at a lower Reynolds number, corresponding to a turbulence intensity of $14.7\%$ at the \myhl{inlet} of the main computational domain. The physical representativeness of the turbulent fluctuations is assessed similarly to the previous section.

\begin{figure}[h!t]
    \centering
    \includegraphics[width=0.7\columnwidth]{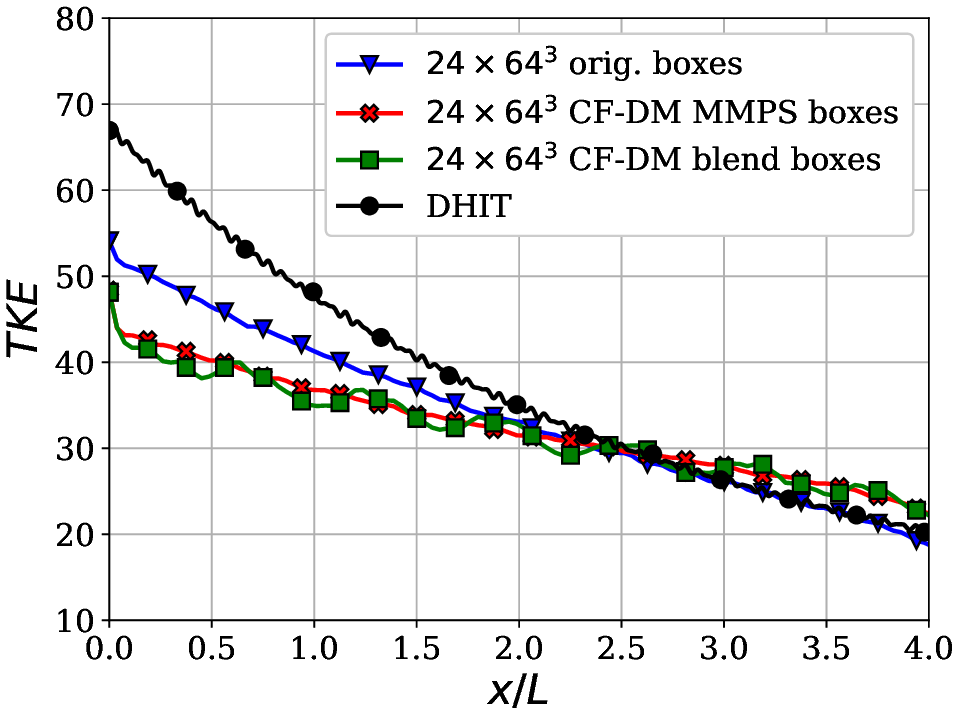}
    \caption{\textbf{Freestream turbulence injection at TI$\bm{=14.7\%}$ to \textit{a posteriori} assess the CF-DM conditioned with $\bm{Re_{L_{int}}=571}$.} Streamwise evolution of the TKE.}
    \label{fig:multi:1500:post:TKE}
\end{figure}

The evolution of the TKE, in Figure~\ref{fig:multi:1500:post:TKE}, for the MMPS and blended boxes takes a longer distance (about two box lengths) to match that of the DHIT, as a larger gap is observed at the inlet due to the underestimation of the inertial range (Figure~\ref{fig:multi:priori:inside:spectrum}). However, the fluctuations imposed by the original precursor method also require two box lengths. A better match of the TKE level in the generated boxes could help closing the gap between the curves.

\begin{figure}[h!t]
    \centering
    \includegraphics[width=0.7\columnwidth]{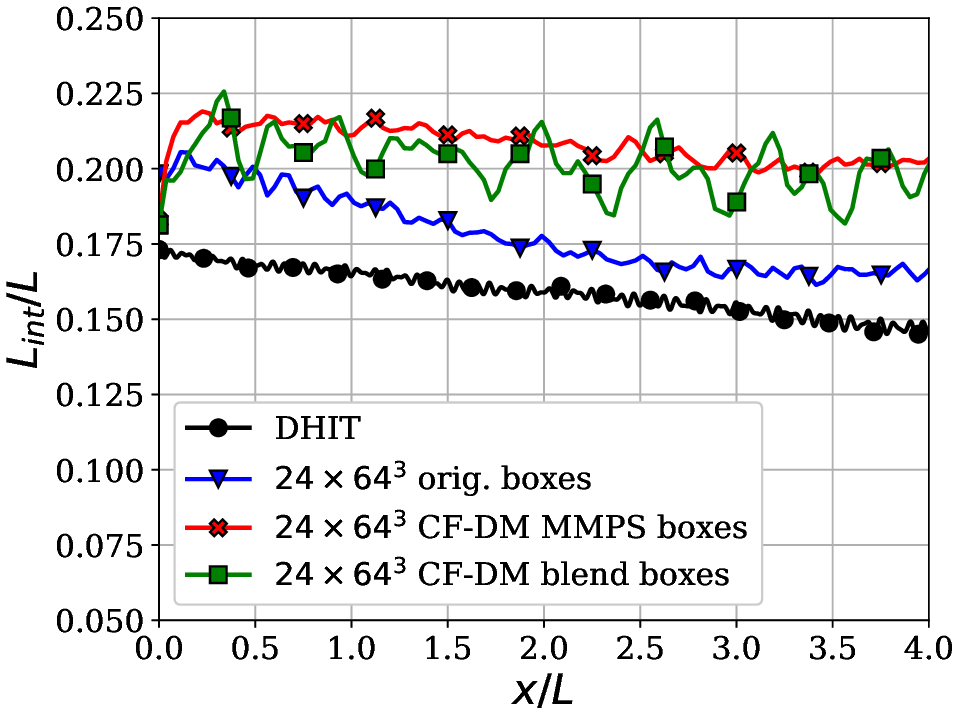}
    \caption{\textbf{Freestream turbulence injection at TI$\bm{=14.7\%}$ to \textit{a posteriori} assess the CF-DM conditioned with $\bm{Re_{L_{int}}=571}$.} Streamwise evolution of the longitudinal length scale.}
    \label{fig:multi:1500:post:ILgth}
\end{figure}

Figure~\ref{fig:multi:1500:post:ILgth} shows the evolution of the integral length scale. Similar to the previous observation, the CF-DM predicts almost the same $L_{int}$ at each TKE level (about $20\%$ of relative error at $x/L=4$). The original precursor method also has difficulty matching the integral length scale in the free domain. This quantity is known sensitive, and very challenging to match with experiments.

\begin{figure*}[h!t]
    \centering
    \includegraphics[width=0.66\linewidth, trim=0 220 430 0, clip=true]{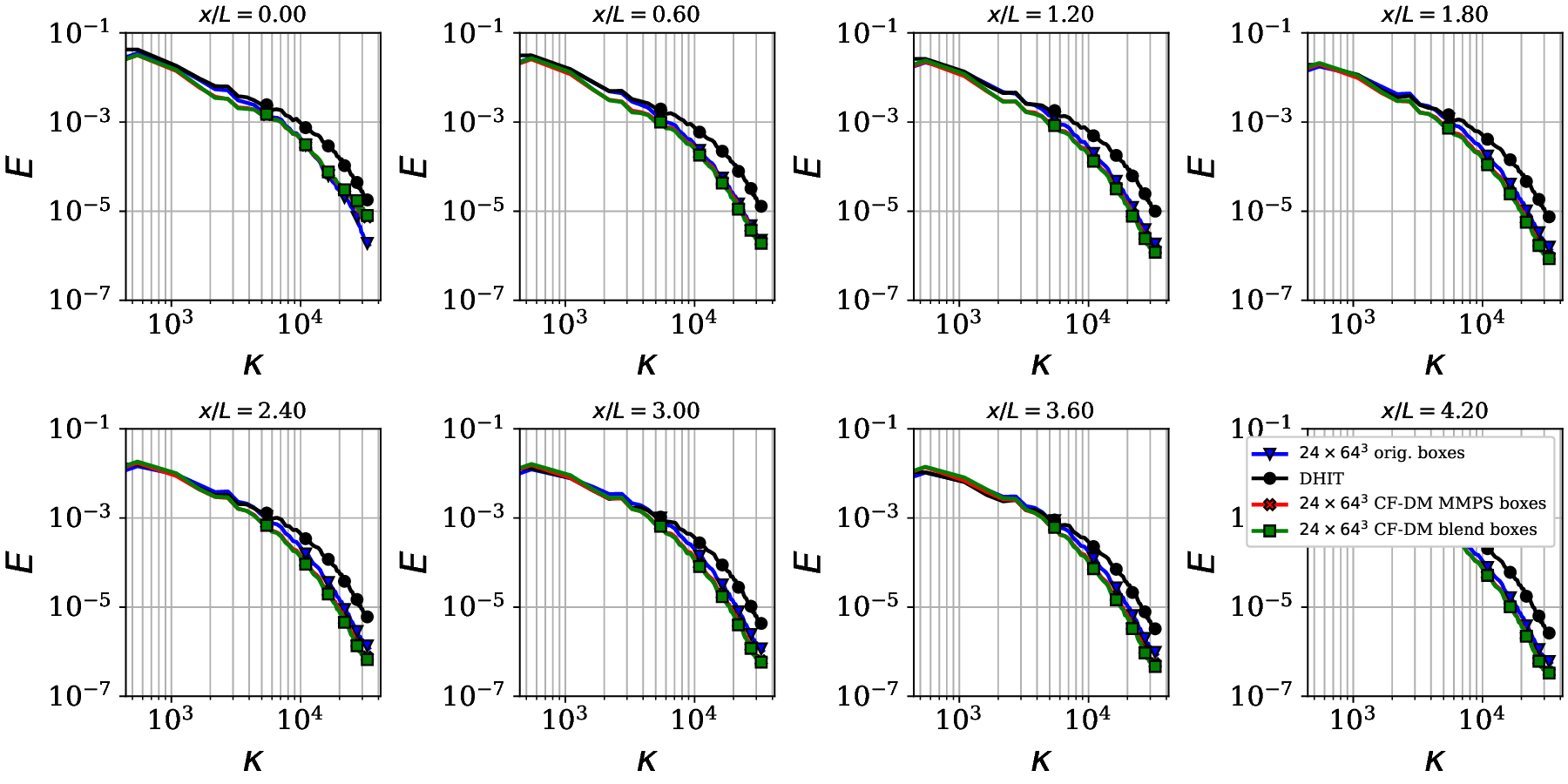}
    \includegraphics[width=0.33\linewidth, trim=0 0 645 220, clip=true]{Figures/ResMultiBox/1500/spectrum.eps}
    \caption{\textbf{Freestream turbulence injection at TI$\bm{=14.7\%}$ to \textit{a posteriori} assess the CF-DM conditioned with $\bm{Re_{L_{int}}=571}$.} 2D energy spectrum at three streamwise locations $x/L=\{0,0.6,2.4\}$ in the free domain, the curves colors follow the color mapping of Figure~\ref{fig:multi:1500:post:TKE}.}
    \label{fig:multi:1500:post:spectrum}
\end{figure*}

Regarding the two-dimensional energy spectrum (Figure~\ref{fig:multi:1500:post:spectrum}), the gap decreases in the inertial range, and a good agreement with the blue triangle curve is observed after two box lengths. 

\begin{figure}[h!t]
    \centering
    \includegraphics[width=0.7\columnwidth]{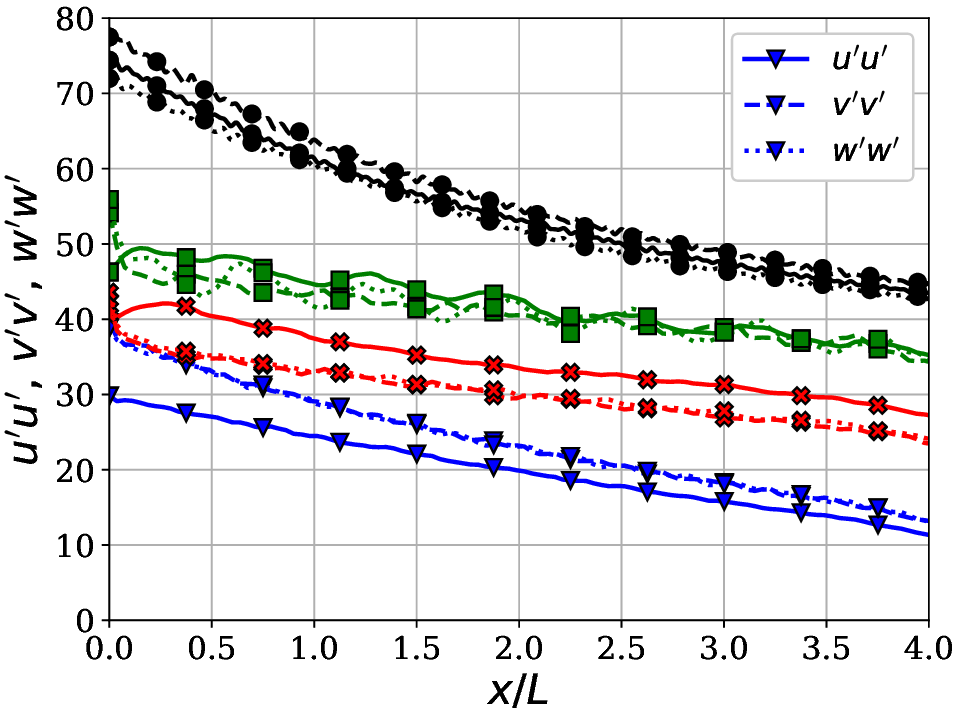}
    \caption{\textbf{Freestream turbulence injection at TI$\bm{=14.7\%}$ to \textit{a posteriori} assess the CF-DM conditioned with $\bm{571}$.} Streamwise evolution of the Reynolds stress components in the free domain, the different sets of curves are shifted by $10$ for readability, the color mapping is similar to Figure~\ref{fig:multi:1500:post:TKE}.}
    \label{fig:multi:1500:post:anisotropy}
\end{figure}

Finally, the evolution of the diagonal components of the Reynolds stress tensor is observed in Figure~\ref{fig:multi:1500:post:anisotropy}. A similar conclusion can be drawn as in Section~\ref{subsubsec:multi:post:850}. The MMPS generated boxes show anisotropy with the $u^\prime u^\prime$ component being dominant. The blended boxes exhibit no anisotropy except at the inlet plane. Nevertheless, the precursor method injects anisotropy in the main computational domain, but this anisotropy reduces with distance from the inlet.


% ---
% Conclusion and perspective (wake, different DHIT box size)
% ---
\section{\label{sec:discuss}Discussion}

Table~\ref{tab:discussion} compares the two strategies for injecting realistic freestream turbulence based on the constraints mentioned by Dhamankar~\textit{et al.}~\citep{DBL2015} regarding the realism of the method, both \textit{a priori} and \textit{a posteriori}, the development length (i.e., the space required for the fluctuations to become fully turbulent), storage, computational cost, and the need to read the inflow data.

Concerning the realism, no method can beat the precursor method \textit{a priori}, that is why the DHIT is used as database in the present work. As shown in Section~\ref{sec:res}, the DM can produce realistic boxes of turbulence at multiple $Re_{L_{int}}$. The vorticity probability distribution of the samples presents similar extreme events as the ground truth. Nonetheless, the model has more difficulties to accurately predict the longitudinal and transverse autocorrelation functions. 

\textit{A posteriori}, the physical representativeness of the \textit{synthetic} turbulent fluctuations is assessed w.r.t. the original precursor method, and a good agreement is observed for all the measured statistics, except for the Reynolds stress tensor, where anisotropy was detected for MMPS generated fluctuations.

\begin{table}[ht]
    \centering
    \caption{\label{tab:discussion}Comparison between the original precursor method and the novel turbulent injection method assisted by CF-DM.}
    \begin{tabular}{ccccc} 
        \hline 
        & & \textbf{Precursor} & \multicolumn{2}{c}{\textbf{CF-DM assisted injection}} \\
        \cline{4-5}
        & &                     & \textbf{Blending} & \textbf{MMPS} \\
        \hline
        & \textbf{Realism} &  &  \\
        \multirow{4}{1em}{\rotatebox{90}{\textit{A priori}}} &  $E(\kappa)$ & \cmark & \multicolumn{2}{c}{\cmark} \\ 
        & $f(r),g(r)$ & \cmark & \multicolumn{2}{c}{$\approx$} \\ 
        & $\triangle$ & \cmark & \multicolumn{2}{c}{\cmark} \\ 
        & $p(\omega_x)$ & \cmark & \multicolumn{2}{c}{\cmark} \\ 
        \cline{2-5}
        \multirow{4}{1em}{\rotatebox{90}{\textit{A post.}}} &  $E(\kappa;x/L)$ & \cmark & \cmark & \cmark \\ 
        & $k(x/L)$ & \cmark & \cmark & \cmark \\ 
        & $L_{int}(x/L)$ & \cmark & $\approx$ & $\approx$ \\ 
        & $\mbox{RST}(x/L)$ & $\approx$ & \cmark &  \xmark \\ 
        \hline 
        & \textbf{Development} & 1 to 2 $L$ & \multicolumn{2}{c}{1 to 2 $L$} \\
        \hline 
        & \textbf{Storage} & $24\times3\times64^3$ & \multicolumn{2}{c}{$32$MB (GPUs)} \\
        & & $\approx 72$MB (CPUs) &  \multicolumn{2}{c}{$+$ $3$MB (CPUs)}  \\
        \hline 
        & \textbf{Computatio-} & $256$ CPU hours & Sampling 24 & MMPS of 1\\
        & \textbf{nal cost} & on Lucia & for boxes in $180$ s & box in $16$ s\\
        \hline 
        & \textbf{IO} & Yes &  \multicolumn{2}{c}{No (comm.} \\
        &  &  &  \multicolumn{2}{c}{GPU $\rightarrow$ CPU)} \\
        \hline
    \end{tabular}
\end{table}

\begin{figure}[h!t]
    \centering
    \includegraphics[width=.7\linewidth]{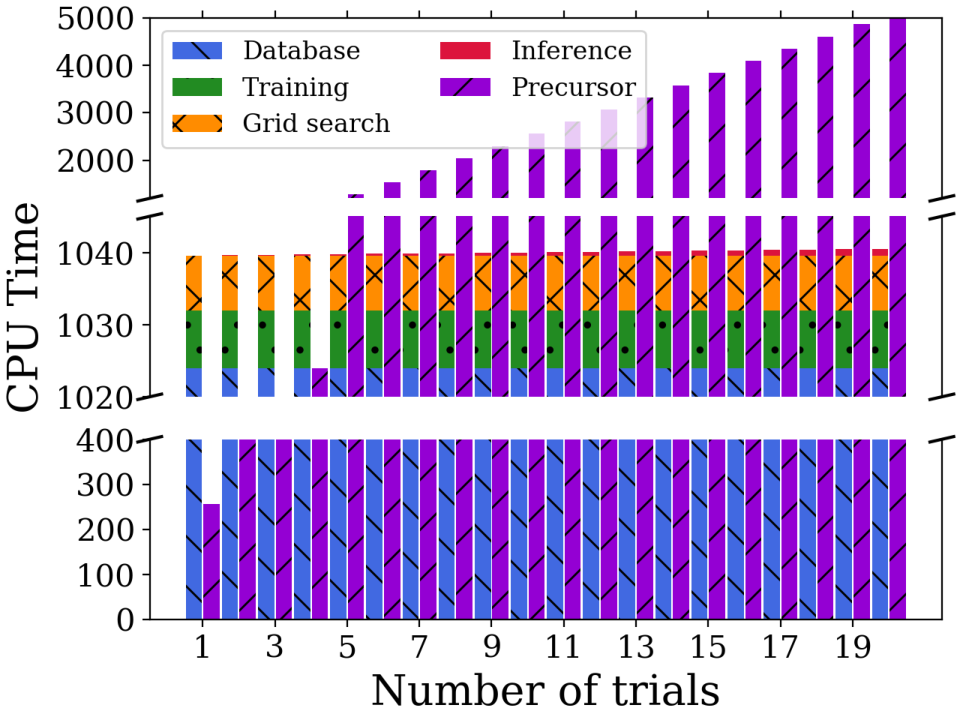}
    \caption{Efficiency graph}
    \label{fig:efficiency}
\end{figure}

The development distance of the two methods is approximately the same, but strongly depends on the injected level of turbulence intensity.

The most significant difference between the two methods lies in their algorithmic performances (as shown in Table~\ref{tab:discussion}). Due to the generative capabilities of DM, new boxes can be generated over and over again without the need to pre-process a single DHIT using rotations and translations to mimic new realizations of turbulence. Therefore, no sample is stored, the only required storage is for the model, which occupies about $32$ MB. This amount of memory may seem insignificant on a modern cluster, but we need to think about the memory left on the CPU nodes by large simulations. For the precursor method, the fluctuations are stored and the boundary condition has to read the inflow from the storage system, while for the data-driven approach, the boxes are sampled on the GPU and then fetched to the CPUs. There is no IO, but communication is required.

\myhl{Figure~\ref{fig:efficiency} shows the efficiency of the ML approach compared to the standard precursor method in CPU time. The ML approach has an initial offset linked to the generation of the database, the training time, and the grid search (as discussed in Section~\ref{subsec:apriori}). Nonetheless, after five trials, the CPU time required to run the DHIT exceeds the initial offset of the ML approach.} The power of CF-DM is significant when it comes to generating samples at different levels of TKE. The model can output 24 boxes in 3 minutes compared to simulating the DHIT which takes about 6144 CPU hours on a modern cluster. A DHIT is not that expensive compared to more realistic configurations. 

The benefits of the ML approach can be leveraged regarding the turnaround time. Using the standard precursor method, a DHIT simulation is first run to extract a given time step. Then, the box is injected into the main computational domain, and the level of TKE is monitored at a given location. If the level is incorrect, another DHIT must be run and frozen at a different TKE level before being injected into the main computational domain. These steps are performed iteratively, which drastically increases the turnaround time. With the data-driven model, training is performed directly on a large collection of frozen DHITs at multiple TKE levels. When using the model within the complete loop, the inference time is very small (see Figure~\ref{fig:efficiency}), and any TKE level within the training range can be queried. Therefore, the turnaround time is drastically reduced.

The proposed method using CF-DM generates reliable turbulent fluctuations while avoiding any storage and accelerating the setup process of finding the appropriate input parameters to obtain the desired level of TKE at the target location in the main computational domain.

\section{\label{sec:conclu}Conclusion and perspectives}

The present work focuses on the development of a novel turbulence injection method using state-of-the-art diffusion models for the prediction of three-dimensional boxes of turbulence fluctuations of the velocity field. The goal is to accelerate the setup of the precursor method implemented in Argo-DG and to reduce the need for storing the fluctuations. 

A series of four DHIT were simulated and a total of 214 instantaneous velocity fields associated with a pair of TKE level and integral length scale were stored to train the diffusion model. A first DM was trained on a single DHIT box to evaluate the ability of the model to generate a realistic turbulent field. The physical representativeness of the samples was evaluated using four statistics: the energy spectrum, the two-point autocorrelation function, the anisotropic state of the Reynolds stress tensor, and the vorticity distribution. The results show good agreement with the ground truth both \textit{a priori} and \textit{a posteriori}.

Training on a single box is too restrictive for real production applications, where a wider range of TKE levels is targeted to speed up injection setup. Therefore, a classifier-free diffusion model was trained on a uniform distribution of the 214 boxes labeled by a unique Reynolds number $Re_{L_{int}}$.

The model was \textit{a priori} assessed by conditioning the generation at four Reynolds numbers: two within, and two outside the training limits. On the one hand, the model shows good interpolation capabilities. The results are in good agreement with the ground truth, except for a slight underestimation of the inertial range at the lowest $Re_{L_{int}}$. The correlation functions are the most difficult to fit, regardless of the $Re_{L_{int}}$ used to condition the model. On the other hand, the model shows poor extrapolation capabilities for $Re_{L_{int}}$ taken above the upper training limit, but good extrapolation capabilities for $Re_{L_{int}}$ taken below the lower limit. This behavior can be explained by the turbulent cascade principle. As the $Re_{L_{int}}$ decreases, the smallest scales are dissipated and we are left with only the largest scales until no more energy is left in the box. Therefore, for smaller $Re_{L_{int}}$ the model only has to dissipate the smallest scales, while for larger $Re_{L_{int}}$ it has to invent structures that were not seen during the training.

\textit{A posteriori}, two level of TI were injected into a free domain and good agreement was observed with the reference case using the original precursor as discussed in Section~\ref{sec:discuss}. 

A first perspective of this work is to validate the approach on an actual turbomachinery blade cascade, a low-pressure compressor blade, designed by Safran Aero Booster~\citep{PMCHDH2019} and measured at the Von Karman Institut.

A second perspective is to guide the training of the diffusion model to impose physical constraints such as the divergence-free and the anisotropic state of the Reynolds stress tensor. The first constraint is really important to avoid the injection of acoustic waves, while the second could probably reduce the anisotropy observed in Figures~\ref{fig:multi:850:post:anisotropy} and~\ref{fig:multi:1500:post:anisotropy}. 

A third perspective is to use control theory or (deep) reinforcement learning to automate the selection of the conditioning of the diffusion model based on sensors placed in the main computational domain, ensuring the correct TKE level compared to experimental campaigns.

Another perspective is to extend the present method to a wider range of fluids, such as wakes, turbulent channel flows, and turbulent boundary layers. Since these test cases are no longer periodic and homogeneous in all directions, the current architecture would potentially be redesigned.

% ---
% Acknoledgment
% ---
\section*{Acknowledgment}
The ARIAC Project (n°2010235) funded by the Service Public de Wallonie (SPW Recherche) is gratefully acknowledged for funding this research. The present research benefited from computational resources made available on Lucia, the Tier-1 supercomputer of the Walloon Region, infrastructure funded by the Walloon Region under the grant agreement n°1910247. The following article has been submitted to/accepted by Physics of Fluid (AIP Publishing LLC). After publication, it will be available at \url{https://doi.org/10.1063/5.0278541}. 

% ---
% Appendix
% ---

\appendix

\renewcommand{\figurename}{Figure A}
\renewcommand{\tablename}{Table A}
\setcounter{figure}{0} 
\setcounter{table}{0}

\section{\label{subsec:app:DDPM}Diffusion Models} 

This section follows the notation of~\citet{KALA2022} and is an extension of Section~\ref{sec:DDPM}, which provides a more detailed explanation of the parameterization of the forward diffusion process in diffusion models.

The forward diffusion process, that progressively brings noise into the images, is governed by the SDE defined in Equation~\ref{eq:forward}. Because the SDE is linear with respect to $\bm{x}_t$, the perturbation (or transition) kernel from $\bm{x}_0$ to $\bm{x}_t$ has the following general form, 
\begin{equation*}
    p_{0t}(\bm{x}_t \vert \bm{x}_0) = \mathcal{N}\left(\bm{x}_t \,\vert\, s(t)\bm{x}_0, \,s^2(t)\sigma(t)^2\mathbf{I}\right),
\end{equation*}
where $s(t)$ and $\sigma_t^2$ are the scaling and the scheduler, respectively. They can be analytically derived from $f(t)$ and $g(t)$ as follows, 
\begin{equation*}
    s(t) = \exp\left( \int_0^t f(\xi) \mathrm{d}\xi \right)\,\, \mbox{ and } \,\, \sigma(t) = \sqrt{\int_0^t \dfrac{g^2(\xi)}{s^2(\xi)} \mathrm{d}\xi}.
\end{equation*}
By integrating the perturbation kernel over $\bm{x}_0$, the marginal distribution $p_t(\bm{x})$ is obtained as,
\begin{equation*}
    p_t(\bm{x}) = \int p_{0t}(\bm{x} \vert \bm{x}_0) p_{\mbox{data}}(\bm{x}_0) \mathrm{d} \bm{x}_0,
\end{equation*}
where $p_{\mbox{data}}(\bm{x})$ denotes the data distribution.

\begin{figure}[h!t]
    \centering
    \includegraphics[width=.45\linewidth]{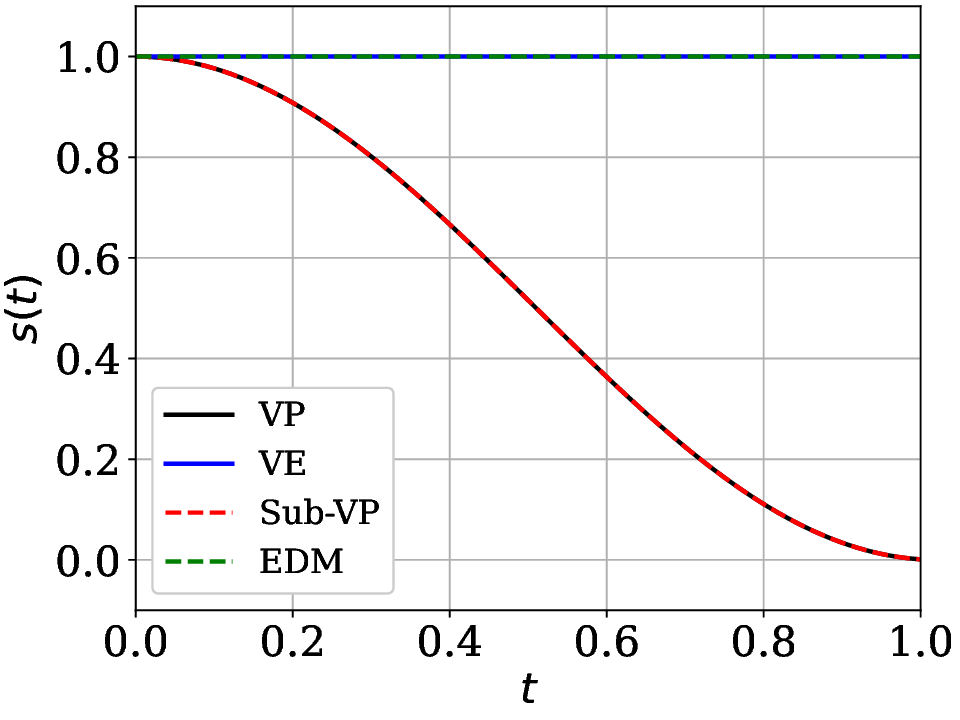}
    \includegraphics[width=.45\linewidth]{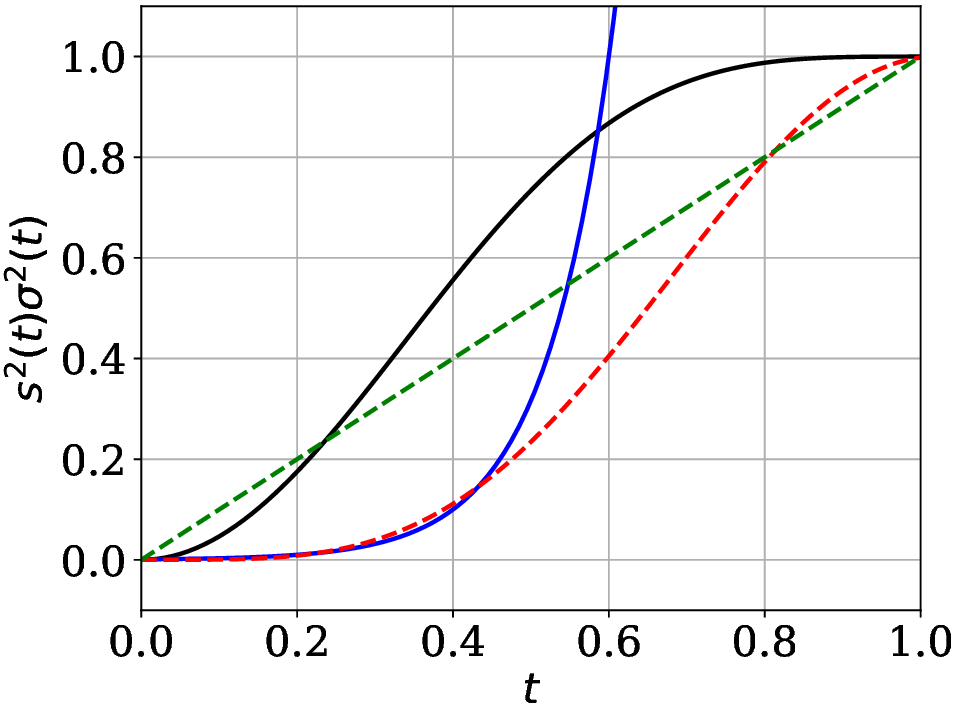}
    \caption{Evolution of the scaling $s(t)$ and the scheduler $\sigma^2(t)$ for the four SDE parametrizations summarized in Table~A~\ref{app:tab:SDE}. }
    \label{app:fig:SDE:scaling}
\end{figure}

There are various way to defined the scaling $s(t)$ and the scheduler $\sigma_t^2$ to ensure that the noise introduced by the forward diffusion process is sufficiently large to ensure that $p_1$ does not depend on $p_{\mbox{data}}$. In other words, the noise introduced by the forward diffusion process must be sufficiently large to ensure that $p_1$ does not depend on $p_{\mbox{data}}$. The prior distribution $p_1$ must also be tractable and easy to sample from. The Variance Exploding (VE), Variance Preserving (VP), and sub-VP SDEs~\citep{HJA2020,YSDKKEP2021} are widely used examples that satisfy these constraints. \citet{KALA2022} also proposed Elucidate Diffusion Model (EDM), a novel parametrization of the SDE. Table~A~\ref{app:tab:SDE} summarizes these four SDE parameterizations and Figure~A~\ref{app:fig:SDE:scaling} shows the evolution of the scaling and the scheduler for these different SDE parametrizations. 

\begin{table*}[h!t]
    \centering
    \caption{\label{app:tab:SDE}The table summarizes four known SDEs: (1) the variance preserving SDE (VPSDE), (2) the sub-VPSDE, (3) the variance exploding SDE (VESDE) and the new implementation proposed by Karras \textit{et al.}~\citep{KALA2022}, here denoted as EDM. The parameter $\alpha_t$ is decreasing with time ($t\in[0,1]$) such that $\alpha_0 \approx 1$ and $\alpha_1 \approx 0$, while $\beta_t$ is increasing.}
        \begin{tabular}{l|ccccc}
            \hline 
            & $f(t)$ & $g(t)$ & $s(t)$ & $\sigma^2(t)$ \\
            \hline 
            \textbf{VPSDE} (Ho \textit{et al.})~\citep{HJA2020}     
            & $\dfrac{1}{2} \dfrac{\mathrm{d}}{\mathrm{d}t}\left(\log\alpha_t\right)$
            & $\left(-\dfrac{\mathrm{d}}{\mathrm{d}t}\left(\log\alpha_t\right)\right)^{1/2}$
            & $\sqrt{\alpha_t}$ & $\dfrac{(1-\alpha_t)}{\alpha_t}$  \\
            \textbf{VESDE} (Song \textit{et al.})~\citep{YSDKKEP2021} 
            & $0$        
            & $\left(\dfrac{\mathrm{d}[\beta_t^2]}{\mathrm{d}t}\right)^{1/2}$
            & $1$  & $\beta_t^2$ \\
            \textbf{Sub-VPSDE} (Song \textit{et al.})~\citep{YSDKKEP2021}
            & $\dfrac{1}{2} \dfrac{\mathrm{d}}{\mathrm{d}t}\left(\log\alpha_t\right)$
            & $\left(-\dfrac{\mathrm{d}}{\mathrm{d}t}\left(\log\alpha_t\right)\left(1-\alpha_t^2\right)\right)^{1/2}$
            & $\sqrt{\alpha_t}$ & $\dfrac{(\alpha_t^2-1)^2}{\alpha_t}$ \\
            \textbf{EDM} (Karras \textit{et al.})~\citep{KALA2022}      
            & $0$       
            & $\left(2t\right)^{1/4}$
            & $1$ & $t$ \\
            \hline 
    \end{tabular}
\end{table*}

As proven by~\citet{YSDKKEP2021}, the resulting SDE obtained with sub-VP performs numerically well in stability and likelihood computations, since the variance of the stochastic process is always bounded by the VPSDE at each intermediate time step (Figure~A~\ref{app:fig:SDE:scaling}). In the present work, the standard VPSDE is adopted to schedule the noise, although other schedulers could have been considered.

\section{\label{subsec:app:post:res}Extension of the \textit{a posteriori} assessment of methodology on multiple boxes}

\begin{figure}
    \begin{minipage}[t]{0.49\linewidth}
         \centering
        \includegraphics[width=\linewidth]{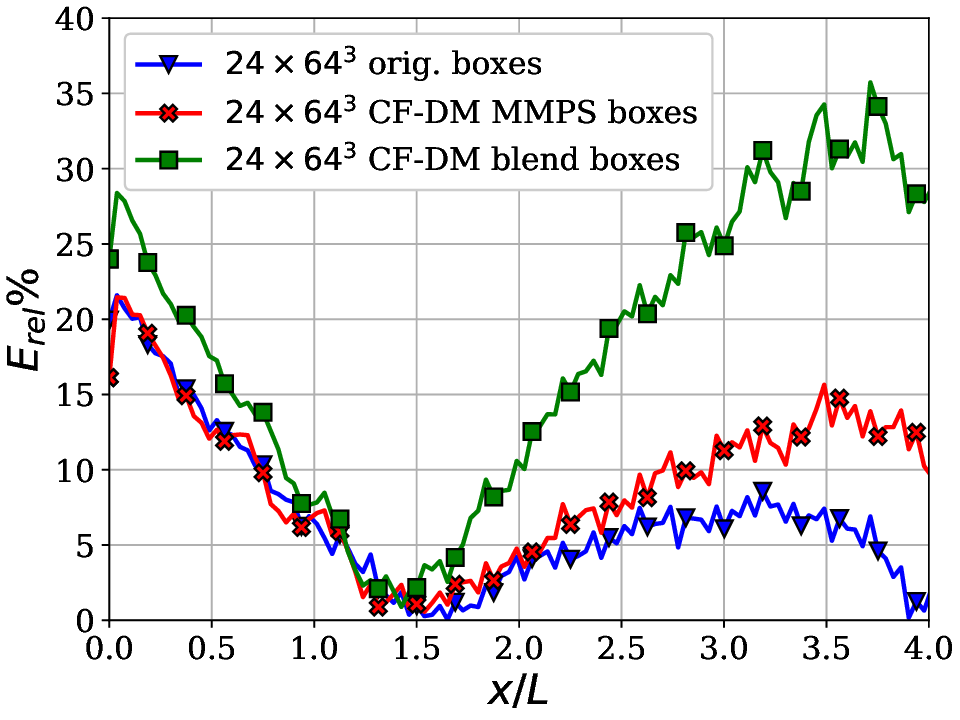}
        \caption{\label{app:fig:multi:850:post:rel}\textbf{Freestream turbulence injection at TI$\bm{=23.3\%}$ to \textit{a posteriori} assess the CF-DM conditioned with $\bm{Re_{L_{int}}=834}$.} Streamwise evolution of the relative error between the TKE level of the DHIT and the level obtained using the precursor method and the ML approach.}
    \end{minipage}
    \hfill
    \begin{minipage}[t]{0.49\linewidth}
        \centering
        \includegraphics[width=\linewidth]{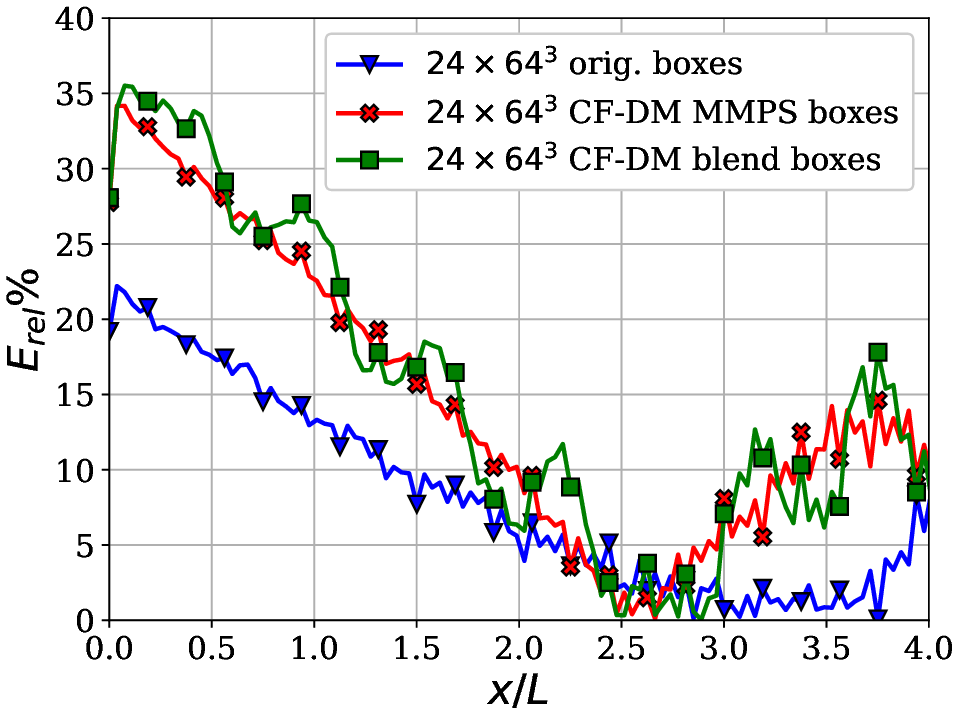}
        \caption{\label{app:fig:multi:1500:post:rel}\textbf{Freestream turbulence injection at TI$\bm{=14.7\%}$ to \textit{a posteriori} assess the CF-DM conditioned with $\bm{Re_{L_{int}}=571}$.} Streamwise evolution of the relative error between the TKE level of the DHIT and the level obtained using the precursor method and the ML approach.}
    \end{minipage}
\end{figure}

Figures A~\ref{app:fig:multi:850:post:rel} and~\ref{app:fig:multi:1500:post:rel} show the relative error between the TKE level of the DHIT (obtained by converting the time to space using the inlet mean velocity) and the level obtained using the standard precursor and the ML approach.

Figure A~\ref{app:fig:multi:850:post:rel} clearly shows that the injection with the blended box (green curve) overestimates the TKE level. Its evolution in the main computational domain differs from that of the DHIT due to the underestimation of the TKE level at the inlet plane. However, the injection using the MMPS approach (red curve) yields results similar to those of the standard precursor approach.

At a lower level of turbulence injection, the diffusion model predicts a slightly lower TKE level compared to the ground truth. This slight error, combined with the strong dissipation observed at the inlet plane, leads to different TKE level evolution in the first part of the domain $(x/L < 2)$, as shown in Figure A~\ref{app:fig:multi:1500:post:rel}. The important location is the matching location at $x/L = 4$, where both the precursor method and the ML method give a similar relative error compared to the expected level of TKE obtained with DHIT. In this case, the user will need to rerun a DHIT simulation to inject a different level of TKE at the inlet and ensure a better match at the matching location. This step takes only a few seconds with the diffusion model.

\begin{figure*}
    \centering
    \includegraphics[width=.7\linewidth]{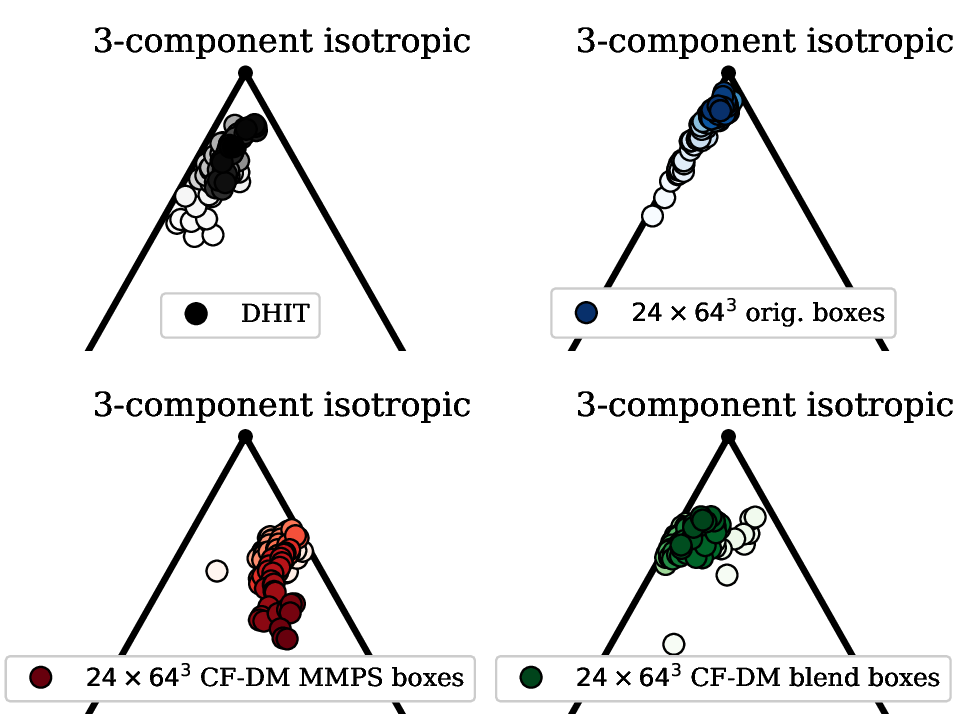}
    \caption{\textbf{Freestream turbulence injection at TI$\bm{=23.3\%}$ to \textit{a posteriori} assess the CF-DM conditioned with $\bm{Re_{L_{int}}=834}$.} Streamwise evolution of the anisotropic state of the Reynolds stress tensor in the main computational domain, light to dark color for the evolution from $x/l=0$ to $x/L=4$. }
    \label{app:fig:multi:850:post:bary}
\end{figure*}

Figure~A~\ref{app:fig:multi:850:post:bary} shows the streamwise evolution of the anisotropic state of the Reynolds stress tensor inside the Barycentric triangle, where the light to dark colors indicate the progression from $x/L=0$ to $x/L=4$. For the original precursor method, the states of the Reynolds stress tensor follow the left edge of the triangle up to the top corner, indicating an isotropic state. For the velocity fluctuations generated with MMPS, the points remain on the right and then evolve towards the bottom of the triangle. This observation is coherent with the anisotropy showed in Figure~\ref{fig:multi:850:post:anisotropy}. For the blended boxes, the states are first located to the right and then converge to the left edge of the triangle. The points are much more clustered compared to the boxes obtained with the original method. 

\begin{figure*}
    \centering
    \includegraphics[width=.7\linewidth]{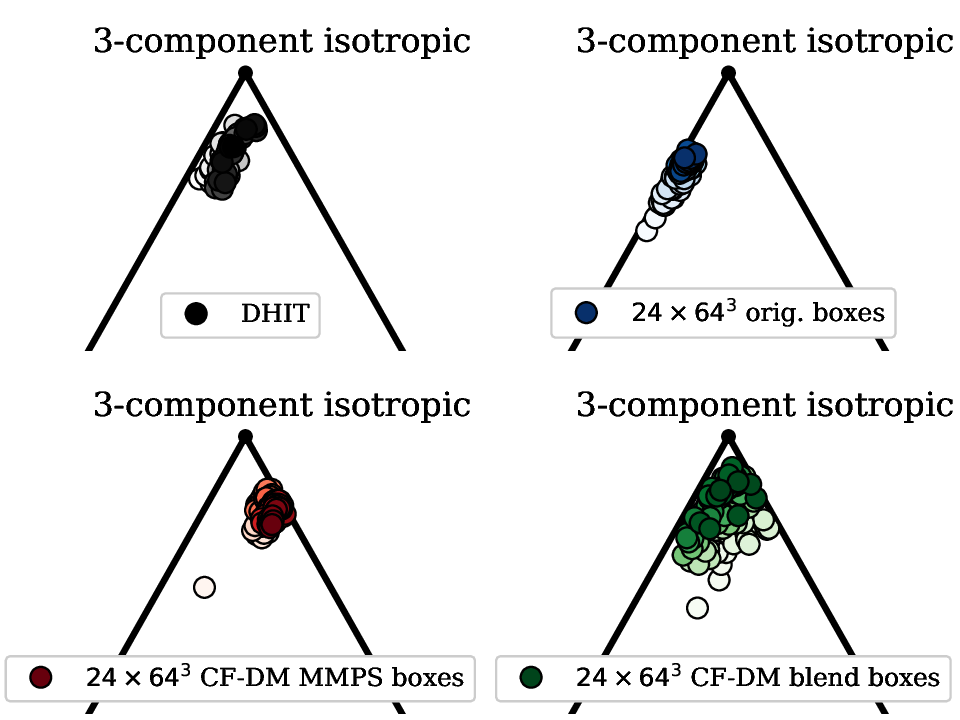}
    \caption{\textbf{Freestream turbulence injection at TI$\bm{=14.7\%}$ to \textit{a posteriori} assess the CF-DM conditioned with $\bm{Re_{L_{int}}=571}$.} Streamwise evolution of the anisotropic state of the Reynolds stress tensor in the main computational domain, light to dark color for the evolution from $x/l=0$ to $x/L=4$. }
    \label{app:fig:multi:1500:post:bary}
\end{figure*}

Figure~A~\ref{app:fig:multi:1500:post:bary} also shows the streamwise evolution of the anisotropic state of the Reynolds stress tensor but for the lower Reynolds number ($Re_{L_{int}}=571$). Concerning the original precursor method, the trend is very similar to in Figure~A~\ref{app:fig:multi:850:post:bary}, except that the points do not meet the upper corner, indicating a slight anisotropy at $x/L=4$, as also observed in Figure~\ref{fig:multi:850:post:anisotropy}. For the velocity fluctuations generated using MMPS, the points are clustered along the right edge of the triangle rather than the left. In the case of the blended boxes, the points do not follow a distinct path but are scattered between the left and right edges of the triangle. This behavior can be attributed to the oscillatory nature of the diagonal terms of the Reynolds stress tensor (see Figure~\ref{fig:multi:850:post:anisotropy}). As approaching $x/L=4$, the points appear to get progressively closer to the upper corner.

% ---
% Produces the bibliography via BibTeX.
% ---
\newpage
\bibliographystyle{unsrtnat}
\bibliography{references}  

\end{document}